\documentclass{iopart}       
\usepackage{iopams}
\usepackage{setstack}

\begin{document}
\title{Semiclassical analysis of Wigner $3j$-symbol}
\author{Vincenzo Aquilanti}
\address{Dipartimento di Chimica, Universit\`a di Perugia, Perugia, 
Italy 06100}
\author{Hal M. Haggard, Robert G. Littlejohn and Liang Yu}
\address{Department of Physics, University of
California, Berkeley, California 94720 USA}

\ead{robert@wigner.berkeley.edu}

\begin{abstract}

We analyze the asymptotics of the Wigner $3j$-symbol as a matrix
element connecting eigenfunctions of a pair of integrable systems,
obtained by lifting the problem of the addition of angular momenta
into the space of Schwinger's oscillators.  A novel element is the
appearance of compact Lagrangian manifolds that are not tori, due to
the fact that the observables defining the quantum states are
noncommuting.  These manifolds can be quantized by generalized
Bohr-Sommerfeld rules and yield all the correct quantum numbers.  The
geometry of the classical angular momentum vectors emerges in a clear
manner.  Efficient methods for computing amplitude determinants in
terms of Poisson brackets are developed and illustrated.

\end{abstract}



\section{Introduction}

This article is a study of the asymptotics of the Wigner $3j$-symbol
from the standpoint of semiclassical mechanics, that is, essentially
multidimensional WKB theory for integrable systems.  The principal
result itself, the leading asymptotic expression for the $3j$-symbol,
has been known since Ponzano and Regge (1968).  Nevertheless our
analysis presents several novel features.  One is the exploration of
Lagrangian manifolds in phase space that are not tori (the usual case
for eigenstates of integrable systems).  Instead, one of the states
entering into the $3j$-symbol is supported semiclassically on a
Lagrangian manifold that is a nontrivial 3-torus bundle over $SO(3)$.
This manifold can be quantized by generalized Bohr-Sommerfeld rules,
whereupon it yields the exact eigenvalues required by the quantum
$3j$-symbol, as well as the correct amplitude and phase of its
asymptotic form.  This unusual Lagrangian manifold arises because the
quantum state in question is an eigenstate of a set of noncommuting
operators.  Other novel features include the expression of the asymptotic
phase of the $3j$-symbol in terms of the phases of Schwinger's
harmonic oscillators and the determination of stationary phase points
by geometrically transparent operations on angular momentum vectors in
three-dimensional space.  Yet another is the representation of
multidimensional amplitude determinants as matrices of Poisson
brackets.  Representations of this type have been known for some time,
but they are generalized here to the case of sets of noncommuting
operators.  The final result is a one-line derivation of the amplitude
of the asymptotic form of the $3j$-symbol.  Similarly brief
derivations are possible for the amplitudes of the $6j$- and
$9j$-symbols.  

In addition our analysis of the $3j$-symbol may prove to be useful for
the asymptotic study of the $3nj$-symbols for higher $n$.  The leading
order asymptotics of the $6j$-symbol were derived by Ponzano and Regge
(1968), but the understanding of the asymptotics of the $9j$-symbol is
still incomplete.  These symbols are important in many applications in
atomic, molecular and nuclear physics, for example, the $9j$-symbols
are needed in atomic physics to convert from an $LS$-coupled basis to
a $jj$-coupled basis.  These symbols are all examples of closed spin
networks, of which more elaborate examples occur in applications, each
of which presents a challenge to asymptotic analysis.  Moreover in
recent years new interest in this subject has arisen from researches
into quantum computing (Marzuoli and Rasetti, 2005) and quantum
gravity, where new derivations of the asymptotics of the Wigner
$6j$-symbol have been produced as well as generalizations to other
groups such as the Lorentz group.  The $3nj$-symbols and their
asymptotics have also been used recently in algorithms for molecular
quantum mechanics (De~Fazio \etal\ 2003 and Anderson and Aquilanti,
2006), which exploit the connections with the theory of discrete
orthogonal polynomials (Aquilanti \etal\ 1995, 2001a,b and references
therein).

The asymptotic formula for the $3j$-symbol is closely related to that
for the $6j$-symbol, being a limiting case of the latter.  These were
first derived by Ponzano and Regge (1968), using intuitive methods and
building on Wigner's earlier result for the amplitude of the
$6j$-symbol (Wigner, 1959).  Later Neville (1971) analyzed the
asymptotics of the $3j$- and $6j$- symbols by a discrete version of
WKB theory, applied to the recursion relations satisfied by those
symbols, without apparently knowing of the work of Ponzano and Regge.
His formulas are not presented in a particularly transparent or
geometrical manner, but appear to reproduce some of the results of
Ponzano and Regge.  The formula for the $3j$-symbol (in the form of
Clebsch-Gordan coefficients) was later derived again by Miller (1974),
who presented it as an example of his general theory of semiclassical
matrix elements of integrable systems.  Miller called on the fact that
the phase of the semiclassical matrix element is a generating function
of a canonical transformation, and used the classical transformation
that most obviously corresponds to the quantum addition of angular
momenta to reconstruct the generating function.  The method leads to a
difficult integral, which, once done, yields the five terms in the
phase of the asymptotic formula for the $3j$-symbol.  Somewhat later
Schulten and Gordon (1975a,b) presented a rigorous derivation of the
Ponzano and Regge results for the $3j$- and $6j$-symbols, using
methods similar to those of Neville but carrying them out in a more
thorough and elegant manner.  Schulten and Gordon also provided
uniform approximations for the transition from the classical to
nonclassical regimes, work that has recently been reanalyzed (Geronimo
\etal, 2004) and extended to non-Euclidean and quantum groups (Taylor
and Woodward, 2005).  Somewhat later Biedenharn and Louck (1981b)
presented a review and commentary of the results of Ponzano and Regge,
as well as a proof based on showing that the result satisfies
asymptotically a set of defining relations for the $6j$-symbol.  More
recently the asymptotics of the $3j$-symbol was derived again by
Reinsch and Morehead (1999), working with an integral representation
constructed out of Wigner's single-index sum for the Clebsch-Gordan
coefficients.  About the same time, Roberts (1999) derived the Ponzano
and Regge results for the $6j$-symbol, using methods of geometric
quantization.  Finally, Freidel and Louapre (2003) presented a
derivation of the asymptotic expression for the square of the
$6j$-symbol, based on an analysis of an $SU(2)$ path integral.  This
work was part of a larger study of generalizations of the $6j$-symbol
to other groups (for example, the $10j$-symbol) that are important in
quantum gravity.  See also Barrett and Steele (2003) and Baez,
Christensen and Egan (2002).

There are many variations on the calculation of the asymptotic forms
of the $3nj$-symbols that have been considered by different authors.
There are asymptotic forms inside and outside the classically allowed
regions; uniform approximations connecting two or more of these
regions; asymptotic forms when only some of the quantum numbers are
large and others small; and higher order terms.  Ponzano and Regge
(1968) covered many of these issues, while Reinsch and Morehead
computed some higher order terms. 

The outline of this paper is as follows.  In Sec.~\ref{scwfis} we
review the semiclassical mechanics of integrable systems in the
generic case that one has sets of commuting observables, drawing
attention to an expression for the amplitude determinant in terms of
Poisson brackets.  In Sec.~\ref{Schwingermodel} we review the
Schwinger model for representing angular momentum operators in terms
of harmonic oscillators.  This model allows us to express angular
momentum eigenstates in terms of wave functions on ${\mathbb R}^n$,
which we use in Sec.~\ref{3jSchwinger} to express the $3j$-symbols in
terms of scalar products of such functions.  In Sec.~\ref{cmSchwinger}
we study the Schwinger model from a classical standpoint, in which an
important element is the reduction of the Schwinger phase
space (the ``large phase space'') by the torus group $T^3$, producing
the Poisson manifold ${\mathbb R}^3 \times {\mathbb R}^3 \times
{\mathbb R}^3$ (``angular momentum space'') and the reduced phase
space $S^2 \times S^2 \times S^2$ (the ``small phase space'').  In
Secs.~\ref{jmtori} and \ref{Wignermanifold} we study the two
Lagrangian manifolds that support the states whose scalar product is
the $3j$-symbol.  One is a conventional invariant torus (the
``$jm$-torus''), but the other, what we call the ``Wigner manifold,''
is compact and Lagrangian but not a torus.  This manifold supports
Wigner's state of zero total angular momentum that enters into the
definition of the $3j$-symbols.  In Secs.~\ref{intersections} we study
the intersections of the $jm$-torus and the Wigner manifold, which are
the stationary phase points of the $3j$-symbol, and show how these can
be found by elementary geometrical considerations in three-dimensional
space (that is, by rotating angular momentum vectors).  The
intersection of the two manifolds turns out to be a pair of 4-tori.
In Sec.~\ref{actionintegrals} we compute the action integrals along
the respective Lagrangian manifolds to points on the two 4-tori, whose
difference is the Ponzano and Regge phase of the $3j$-symbol.  In
Sec.~\ref{BSquant} we apply generalized Bohr-Sommerfeld quantization
to the $jm$-torus and the Wigner manifold, a standard procedure for
the $jm$-torus, although it leads in an interesting way to the extra
$1/2$ in the classical values representing the lengths of the angular
momentum vectors.  This extra $1/2$ was guessed by Ponzano and Regge
and Miller and derived systematically by Schulten and Gordon, Reinsch
and Morehead and by us, although it is missing from the results of
Roberts.  In our work it is essentially a Maslov index.  In
Sec.~\ref{ampdet} we generalize known expressions for the amplitude
determinant of semiclassical matrix elements of integrable systems in
terms of Poisson brackets to the case of collections of noncommuting
observables (whose level sets nevertheless are Lagrangian).  The
result allows us to compute the amplitude of the $3j$-symbol as a
$2\times 2$ matrix of Poisson brackets.  We then put all the pieces
together to obtain the final asymptotic form.  Finally, in
Sec.~\ref{conclusions} we present some comments on the work, prospects
for further work, and conclusions.

\section{Semiclassical wave functions for integrable systems}
\label{scwfis}

The semiclassical mechanics of integrable systems is well understood
(Einstein, 1917; Brillouin, 1926; Keller, 1958; Percival, 1973; Berry
and Tabor, 1976; Gutzwiller, 1990; Brack and Bhaduri, 1997; Cargo
\etal, 2005a, 2005b).  Here we summarize the basic facts, some of
which require modification for our application.

We consider the quantum mechanics of a particle moving in ${\mathbb
R}^n$ (with wave function $\psi(x_1,\ldots,x_n)$ and Hilbert space
$L^2({\mathbb R}^n)$).  We speak of an integrable system if we have a
complete set of commuting observables $\{{\hat A}_1, \ldots, {\hat
A_n}\}$ acting on this Hilbert space.  We use hats to distinguish
quantum operators from classical quantities with a similar meaning.
Sometimes the Hamiltonian is one of these operators or a function of
them, but in our application there is no Hamiltonian, or, rather all
the ${\hat A}_i$'s are Hamiltonians on an equal footing.  These
operators may be converted into their classical counterparts by the
Weyl transform (Weyl 1927, Wigner 1932, Groenewold 1946, Moyal 1949,
Voros 1977, Berry 1977, Balazs and Jennings 1984, Hillery \etal\ 1984,
Littlejohn 1986, McDonald 1988, Estrada \etal\ 1989, Gracia-Bond\'\i a
and V\'arilly 1995 and Ozorio de Almeida 1998).  The Weyl transforms
(or Weyl ``symbols'') of these operators are functions on the
classical phase space ${\mathbb R}^{2n}$, that is, functions of
$(x_1,\ldots,x_n; p_1,\ldots,p_n)$.  They are normally even power
series in $\hbar$, as we assume, of which the leading term is the
``principal symbol.''  We denote the principal symbols of $\{{\hat
A}_1, \ldots, {\hat A_n}\}$ by $\{A_1, \ldots, A_n\}$ (without the
hats).  In view of the Moyal star product representation (Moyal 1949)
of the vanishing commutators $[{\hat A}_i, {\hat A}_j]=0$, the
principal symbols Poisson commute, $\{A_i, A_j\}=0$, thus defining a
classically integrable system (Arnold 1989, Cushman and Bates 1997).
(We use curly brackets $\{\;\}$ both to denote a set and for Poisson
brackets.)  Then according to the Liouville-Arnold theorem (Arnold
1989), the compact level sets of $\{A_1, \ldots, A_n\}$ are
generically $n$-tori.  The Hamiltonian vector fields generated by the
$A_i$ are commuting and linearly independent on the tori; thus the
tori are not only the level sets of the $A_i$, they are also the
orbits of the Abelian group generated by the corresponding Hamiltonian
flows.  One can define an action function $S$ on a torus as the
integral of $\sum_i p_i \, dx_i$ relative to some initial point; it is
multivalued because of the topologically distinct paths going from the
initial to the final point, but otherwise is independent of the path.

Let $A_i=a_i$ be one of these tori ($A_i$ are the functions, $a_i$ the
values).  The torus has a projection onto configuration space defining
a classically allowed region in that space; the inverse projection is
multivalued.  The function $S$ may be projected onto configuration
space, defining a function we shall denote by $S_k(x,a)$ (where for
brevity $x$ and $a$ stand for $(x_1, \ldots,x_n)$ and $(a_1, \ldots,
a_n)$, etc).  Here $k$ labels the branches of the inverse projection;
function $S_k$ has an additional multivaluedness due to the choice of
contour connecting initial and final points on the torus.  Then as
explained by Arnold (1989), $S(x,a)$ is the generating function of the
canonical transformation $(x,p) \to (\alpha,A)$, where
$\alpha=(\alpha_1, \ldots, \alpha_n)$ is the set of angle variable
conjugate to the conserved quantities $(A_1, \ldots, A_n)$.  Action
variables may be defined in the usual way as $(1/2\pi) \oint p\,dx$
around the independent basis contours on the torus; these are
functions of the $A_i$ and their conjugate variables are the angles
that cover the torus once when varying between $0$ and $2\pi$.
Sometimes however it is more convenient to work with the $A_i$ instead
of the actions (the $A_i$ are not necessarily actions, and their flows
are not necessarily periodic on the torus).

Tori are quantized, that is, associated with a consistent solution of
the simultaneous Hamiltonian-Jacobi and amplitude transport equations
for the operators ${\hat A}_i$, only if they satisfy the
Bohr-Sommerfeld or EBK quantization conditions, discussed in
Sec.~\ref{BSquant}.  Associated with a quantized torus is a
semiclassical wave function in configuration space, which in the
classically allowed region is given by
\begin{equation}
\psi(x) = \langle x \vert a \rangle =
\frac{1}{\sqrt V} \sum_k |\Omega_k|^{1/2} 
\exp\{i[S_k(x,a) - \mu_k\pi/2]\}.
\label{scpsi}
\end{equation}
The meaning of this formula is the following.  First, here and below
we set $\hbar=1$.  Next, given the point $x$ in the classically
allowed region, its inverse projection onto the quantized torus is a
set of points indexed by $k$.  We assume the projection is nonsingular
at these points (we are not at a caustic).  The phase $S_k(x,a)$ is
the integral of $p\,dx$ from a given initial point on the torus to the
$k$-th point of the inverse projection, and $\mu_k$ is the Maslov
index (Maslov 1981, Mishchenko \etal\ 1990, de Gosson 1997) of the
same path.  The amplitude determinant $\Omega_k$ is given by
\begin{equation}
\Omega_k= \det\frac{\partial^2 S_k(x,a)}{\partial x_i \partial a_j}
	=[ \det \{x_i,A_j\} ]^{-1},
\label{ampldet}
\end{equation}
where in the second form the Poisson brackets are evaluated on the
$k$-th branch of the inverse mapping from $x$ to the Lagrangian
manifold.  The amplitude determinant is a density on configuration
space (to within the semiclassical approximation, the probability
density corresponding to a single branch), which is the projection
onto configuration space of a density on the torus.  The latter
density is required to be invariant under the Hamiltonian flows
generated by the $A_i$ (this is the meaning of the amplitude transport
equations for the $A_i$); in terms of the variables $\alpha_i$
conjugate to the $A_i$ this means that the density is constant (it is
the $n$-form $d\alpha_1 \wedge \ldots \wedge d\alpha_n$).  Finally,
the quantity $V$ in (\ref{scpsi}) is the volume of the torus,
measured with respect to this density.  If the $A_i$ are action
variables, then $V=(2\pi)^n$.  The overall phase of the wave function
(its phase convention) is determined by the choice of the initial
point on the torus.

Now let $\{{\hat A}_1, \ldots, {\hat A}_n\}$ and $\{{\hat B}_1,
\ldots, {\hat B}_n\}$ be two complete sets of commuting observables,
with principal symbols $A_i$ and $B_i$, conjugate angles $\alpha_i$
and $\beta_i$ and action functions $S_A(x,a)$ and $S_B(x,b)$, and let
$a$ and $b$ refer to two quantized tori (an $A$-torus and a
$B$-torus).  We assume initially that the two sets of functions $A_i$
and $B_i$ are independent.  We compute $\langle b \vert a \rangle$ 
as an integral of the wave functions over $x$, evaluated by the
stationary phase approximation.  The stationary phase points are
geometrically the intersections of the $A$-torus with the $B$-torus.
Generically the two tori intersect in finite set of isolated points
that we index by $k$, denoting the corresponding $\alpha$ and $\beta$
values by $\alpha_k$ and $\beta_k$.  (For given $k$, $\alpha_k$ and
$\beta_k$ refer to the same point in phase space.)  Then the result is
\begin{equation}
\langle b\vert a\rangle = 
\frac{(2\pi i)^{n/2}}{\sqrt{V_A V_B}}
\sum_k |\Omega_k|^{1/2}
\exp\{ i[S_A(\alpha_k)-S_B(\beta_k)-\mu_k\pi/2]\}.
\label{abmatrixelement}
\end{equation}
Here $V_A$ and $V_B$ are the volumes of the respective tori, as in
(\ref{scpsi}), and the actions $S_A$ and $S_B$ are considered
functions of the $\alpha$ or $\beta$ coordinates on the respective
tori.  

As shown by Littlejohn (1990), the amplitude determinant $\Omega_k$
can be written in terms of the Poisson brackets of the observables
$A_i$ and $B_i$,
\begin{equation}
\Omega_k = [\det \{A_i, B_j\}]^{-1}.
\label{PBdet}
\end{equation}
The Maslov index $\mu_k$ in (\ref{abmatrixelement}) is not the
same as in (\ref{scpsi}).

Another case considered by Littlejohn (1990) is the one in which some
of the $A_i$ are functionally dependent on some of the $B_i$.  For
this case it is convenient to assume that the first $r$ of the two
sets of variables $A$ and $B$ are functionally independent, while the
last $n-r$ are identical, so that $A=\{A_1, \ldots, A_r, A_{r+1},
\ldots, A_n\}$ and $B=\{B_1, \ldots, B_r, A_{r+1}, \ldots, A_n\}$.
Then the stationary phase points are still the intersections of the
two $n$-tori, but now the intersections are generically a finite set
of isolated $(n-r)$-tori, upon which linearly independent vector
fields are the Hamiltonian vector fields associated with the
$(A_{r+1}, \ldots, A_n)$.  Such an $(n-r)$-torus is the orbit of the
Abelian group action generated by the corresponding Hamiltonian flows.
In this case we find
\begin{equation}
\langle b \vert a \rangle =
\frac{(2\pi i)^{r/2}}{\sqrt{V_A V_B}}
\sum_k V_k |\Omega_k|^{1/2} 
\exp\{i[S_A(\alpha_k)-S_B(\beta_k)-\mu_k\pi/2]\},
\label{abmatrixelement1}
\end{equation}
where now $V_k$ is the volume of the $k$-th intersection (an
$(n-r)$-torus on which the volume measure is $d\alpha_{r+1} \wedge
\ldots \wedge d\alpha_n$), and where now the amplitude determinant 
$\Omega_k$ is still given by (\ref{PBdet}), except that it is
understood that only the first $r$ of the $A$'s and $B$'s enter (thus,
it is an $r\times r$ determinant instead of an $n \times n$ one).  The
phase difference $S_A - S_B$ for branch $k$ can be evaluated at any
point on the $(n-r)$-torus which is the intersection, since the
integral of $p\,dx$ back and forth along a path lying in the
intersection vanishes.

\section{The Schwinger model}
\label{Schwingermodel}

The Schwinger (or $SU(2)$ or boson) model for angular momenta is
explained well in Schwinger's original paper (reprinted by Biedenharn
and van Dam (1965), the original 1952 paper being unpublished), and
reworked in an interesting way by Bargmann (1962).  For further
perspective see Biedenharn and Louck (1981a) and Smorodinskii and
Shelepin (1972).  Introductions are given by Sakurai (1994) and
Schulman (1981).  Here we define the notation for the Schwinger model
and emphasize some aspects that will be important for our application.

In the Schwinger model each independent angular momentum vector is
associated with two harmonic oscillators.  We shall refer to the
$1j$-, $3j$-, etc models, depending on how many independent angular
momenta there are.  The number of $j$'s in the model is not
necessarily the number of $j$'s in the Wigner symbol; for example,
Miller (1974) used a $2j$-model to study the Clebsch-Gordan
coefficients, essentially the $3j$-symbols.  

We start with the $1j$-model, for which there are two harmonic
oscillators indexed by Greek indices $\mu, \nu, \ldots = 1,2$.  (These
are just labels of the two oscillators; sometimes other labels such as
$1/2$, $-1/2$ are more suitable.)  The wave functions are $\psi(x_1,x_2)$
and the Hilbert space is ${\cal H} = L^2({\mathbb R}^2)$.  We write
${\hat H}_\mu=(1/2)({\hat x}_\mu^2 + {\hat p}_\mu^2)$ for the two
oscillator Hamiltonians, and we define ${\hat H}=\sum_\mu {\hat
H}_\mu$.  The eigenvalues of ${\hat H}$ are $n+1$, with
$n=0,1,\ldots$, and energy level $E_n$ is $(n+1)$-fold degenerate.  We
introduce usual annihilation and creation operators $a_\mu= ({\hat
x}_\mu + i{\hat p}_\mu)/\sqrt{2}$, $a^\dagger_\mu = ({\hat x}_\mu -i
{\hat p}_\mu)/\sqrt{2}$, omitting the hats on the $a$'s and
$a^\dagger$'s since these will always be understood to be operators.
We define operators
\begin{equation}
  {\hat I} = \frac{1}{2} \sum_\mu a^\dagger_\mu a_\mu 
  =\frac{1}{2} ({\hat H} - 1)
\label{Imudef}
\end{equation}
and
\begin{equation}
  {\hat J}_i = \frac{1}{2} 
  \sum_{\mu\nu} a^\dagger_\mu \, \sigma^i_{\mu\nu} \, a_\nu,
\label{Jidef}
\end{equation}
where $\sigma^i$ is the $i$-th Pauli matrix.  Here and below we use
indices $i,j,\ldots = 1,2,3$ (or $x,y,z$ if that is more clear) to
denote the Cartesian components of a 3-vector.  Notice that ${\hat I}$
and ${\hat J}_i$ are quadratic functions of the $x$'s and $p$'s of the
system.  The eigenvalues of ${\hat I}$ are $n/2$ for $n=0,1,\ldots$.
These operators satisfy the commutation relations $[{\hat I},{\hat
J}_i] =0$ and $[{\hat J}_i, {\hat J}_j]= i\sum_k \epsilon_{ijk} {\hat
J}_k$.  We also define ${\hat{\bf J}}^2 = \sum_i {\hat J}_i^2$, so
that $[{\hat I},{\hat{\bf J}}^2]=0$ and $[{\hat J_i}, {\hat{\bf
J}}^2]=0$.  It avoids some confusion with indices to always denote the
square of a vector by a bold face symbol, as we have done here.  We
note the important operator identity ${\hat{\bf J}}^2 = {\hat I}({\hat
I} +1)$, expressing the quartic operator ${\hat{\bf J}}^2$ as a
function of the quadratic operator ${\hat I}$.  

From this identity and the known eigenvalues of ${\hat I}$ it follows
that the eigenvalues of ${\hat{\bf J}}^2$ are $(n/2)[(n/2)+1]$, for
$n=0,1,\ldots$, which leads us to identify $n/2$ with
$j=0,1/2,1,\ldots$, the usual angular momentum quantum number.  The
$n$-th (or $j$-th) eigenspace of ${\hat H}$ or ${\hat I}$ is
$(2j+1)$-dimensional, and so must contain a single copy of the $j$-th
irrep of $SU(2)$.  Each irrep (both integer and half-integer values of
$j$) occurs precisely once in the Hilbert space ${\cal H}$.  We denote
these subspaces by ${\cal H}_j$, and write ${\cal H} = \sum_j \oplus
{\cal H}_j$.  The standard basis in ${\cal H}_j$ is the eigenbasis of
${\hat J}_z =\frac{1}{2}({\hat H}_1 - {\hat H}_2)$, with the usual
quantum number $m$, so that if $n_\mu$ are the usual quantum numbers
of the oscillators ${\hat H}_\mu$, then $n_1=j+m$, $n_2=j-m$.  The
simultaneous eigenstates of ${\hat{\bf J}}^2$ and ${\hat J}_z$ are
$\vert jm\rangle$ or $\vert n_1 n_2
\rangle$.

In the $Nj$-model we index the angular momenta with indices $r,s,\dots
= 1, \ldots, N$.  The oscillators are now labelled ${\hat H}_{r\mu}$
with coordinates and momenta ${\hat x}_{r\mu}$ and ${\hat p}_{r\mu}$
and annihilation and creation operators $a_{r\mu}$ and
$a^\dagger_{r\mu}$.  The wave functions are now $\psi(x_{11}, x_{12},
x_{21}, \ldots, x_{N2})$ and the Hilbert space is $L^2({\mathbb
R}^{2N})$.  We define operators
\begin{eqnarray}
  {\hat I}_r &= \frac{1}{2} \sum_\mu a^\dagger_{r\mu} a_{r\mu},
  \label{Irdef} \\
  {\hat J}_{ri} &= \frac{1}{2} \sum_{\mu\nu}
   a^\dagger_{r\mu} \, \sigma^i_{\mu\nu} \, a_{r\nu},
   \label{Jridef} \\
   {\hat{\bf J}}^2_r &= \sum_i {\hat J}_{ri}^2,
   \label{Jrsquareddef} \\
   {\hat J}_i &= \sum_r {\hat J}_{ri},
   \qquad \hbox{\rm or} \qquad
   {\hat{\bf J}} = \sum_r {\hat{\bf J}}_r, 
   \label{Jtotalidef} \\
   {\hat{\bf J}}^2 & = \sum_i {\hat J}_i^2,
   \label{Jtotalsquareddef}
\end{eqnarray}
most of which are obvious generalizations from the case $N=1$.  These
satisfy the identity
\begin{equation}
  {\hat{\bf J}}_r^2 = {\hat I}_r({\hat I}_r+1),
  \label{JIidentity}
\end{equation}
and the commutation relations $[{\hat I}_r, {\hat J}_{si}] = [{\hat
I}_r, {\hat{\bf J}}^2_s]=0$.  Each angular momentum vector ${\hat{\bf
J}}_r$ also obeys the standard commutation relations among its
components and square, which we omit, as does any sum of these angular
momenta (partial or total).

The angular momenta generate an action of $[SU(2)]^N$ on the Hilbert
space (one copy for each pair of oscillators).  Here we discuss only
the simultaneous rotation of all oscillator degrees of freedom by the
same element of $SU(2)$, which is generated by the total angular
momentum, but  partial rotation operators can also be defined and are
useful.  We begin with the commutation relations,
\begin{eqnarray}
[{\hat J}_i, a_{r\mu}] &= -\frac{1}{2} \sum_\nu \sigma^i_{\mu\nu} \,
a_{r\nu},
\label{Jacomrel} \\ \relax
[{\hat J}_i, a^\dagger_{r\mu}] &=
+\frac{1}{2} \sum_\nu a^\dagger_{r\nu} \,\sigma^i_{\nu\mu},
\label{Jadaggercomrel}
\end{eqnarray}
which define the transformation properties of the operators
$a_{r\mu}$, $a^\dagger_{r\mu}$ under infinitesimal rotations.  We
define a finite rotation operator in axis-angle or Euler angle form by
\begin{eqnarray}
U({\bf n},\theta) &= \exp(-i\theta {\bf n} \cdot {\bf J}),
\label{Uaxisangledef} \\
U(\alpha,\beta,\gamma) &= U({\bf z},\alpha) U({\bf y},\beta)
U({\bf z},\gamma),
\label{UEulerangledef}
\end{eqnarray}
where ${\bf n}$ is a unit vector defining an axis and $\theta$ an
angle of rotation about that axis, and where ${\bf x}$, ${\bf y}$ and
${\bf z}$ are respectively the unit vectors along the three coordinate
axes.  The $U$ operators form a faithful representation of $SU(2)$. 

We use the symbol $u({\bf n},\theta)$ or $u(\alpha,\beta,\gamma)$
for the $2 \times 2$ matrices belonging to $SU(2)$, in axis-angle or
Euler angle parameterization (not to be confused with the $U$
operators that act on the Hilbert space of the $2N$ oscillators).  Thus
\begin{equation}
u({\bf n},\theta) = 
\exp(-i\theta{\bf n}\cdot \bsigma/2)=
\cos\theta/2 - i{\bf n}\cdot {\bsigma}
\sin\theta/2.
\label{udef}
\end{equation}
The exponentiated versions of equations (\ref{Jacomrel}) and
(\ref{Jadaggercomrel}) are
\begin{equation}
  \eqalign{
    U^\dagger a_{r\mu} U &= \sum_\nu u_{\mu\nu} \, a_{r\nu}, \\
    U^\dagger a^\dagger_{r\mu} U &= \sum_\nu
    a^\dagger_{r\nu}\, (u^{-1})_{\nu\mu},}
  \label{rotatea}
\end{equation}
where both $U$ and $u$ have the same parameterization. In the language
of irreducible tensor operators the pair of operators
$(a_{r1},a_{r2})$ transforms as a spin-$1/2$ operator.

Similarly, vector operators are the angular momenta themselves, which
satisfy the conjugation relations,
\begin{equation}
  U^\dagger {\hat J}_{ri} U = \sum_j R_{ij} {\hat J}_{rj},
  \label{rotateJr}
\end{equation}
where $R$ is the $3\times 3$ orthogonal rotation matrix with the same
axis and angle as $U$.  The relation between $R$ and $u$ (with the
same axis and angle) is
\begin{equation}
  R_{ij} = \frac{1}{2} \mathop{\rm tr}(U^\dagger \sigma_i U \sigma_j).
  \label{Ruproj}
\end{equation}
This is the usual projection from $SU(2)$ to $SO(3)$, in which the
inverse image of a given $R \in SO(3)$ is a pair $(u,-u)$ in $SU(2)$.

\section{The Wigner $3j$-symbols in the Schwinger Model}
\label{3jSchwinger}

We now define the $3j$-symbols in the context of the Schwinger
model.  We take the $3j$-model, $N=3$.  One complete set of commuting
observables on the Hilbert space ${\cal H} \otimes {\cal H}
\otimes {\cal H}$ is $({\hat I}_1, {\hat I}_2, {\hat I}_3, {\hat
J}_{1z}, {\hat J}_{2z}, {\hat J}_{3z})$, with  corresponding eigenstates
$\vert j_1j_2j_3m_1m_2m_3 \rangle = \vert j_1 m_1 \rangle \vert j_2
m_2 \rangle \vert j_3 m_3\rangle$.  Another complete set arises in the
usual problem of addition of three angular momenta, in which we
consider the values of $j$ and $m$ (the quantum numbers of ${\hat{\bf
J}}^2$ and ${\hat J}_z$) that occur in the product space ${\cal
H}_{j_1} \otimes {\cal H}_{j_2} \otimes {\cal H}_{j_3}$ for fixed
values of $(j_1,j_2,j_3)$, a subspace of ${\cal H} \otimes {\cal H}
\otimes {\cal H}$.  The set of the five commuting operators $({\hat
I}_1, {\hat I}_2, {\hat I}_3, {\hat J}_3, {\hat{\bf J}}^2)$ that
arises in this way is however not complete (the simultaneous
eigenstates in general possess degeneracies), so to resolve these we
introduce a sixth commuting operator, conventionally taken to be
${\hat{\bf J}}^2_{12} = ({\hat{\bf J}}_1 + {\hat{\bf J}}_2)^2$ with
quantum number $j_{12}$ (${\hat{\bf J}}_{23}^2$ or ${\hat{\bf
J}}_{13}^2$ will also work).

The Wigner $3j$-symbols only involve the case $j=0$, but we mention
the others anyway because the foliation of the classical phase space
into Lagrangian manifolds involves the other values.  The usual rules
for the addition of angular momenta show that if $(j_1,j_2,j_3)$
satisfy the triangle inequality, then there exists precisely a
one-dimensional subspace of ${\cal H}_{j_1} \otimes {\cal H}_{j_2}
\otimes {\cal H}_{j_3}$ with $j=0$; if they do not, then no such
subspace exists.  If we enlarge our point of view to the full Hilbert
space ${\cal H} \otimes {\cal H} \otimes {\cal H}$, then there is an
infinite dimensional subspace with $j=0$, a basis in which is
specified by all triplets $(j_1,j_2,j_3)$ that satisfy the triangle
inequalities.  If $j=0$, then the quantum number $m$ is superfluous,
since $m=0$; the quantum number $j_{12}$ is superfluous as well, since
$j_{12}=j_3$.  

We note that if $\langle \psi \vert {\hat{\bf J}}^2
\vert \psi \rangle=0$ for any state $\vert \psi \rangle$, then ${\hat
J}_i \vert \psi \rangle=0$ for $i=1,2,3$.  Although the components of
${\hat{\bf J}}$ do not commute and so do not possess simultaneous
eigenstates in general, the case of a state with $j=0$ is an
exception, since it is a simultaneous eigenstate of all three
components of ${\hat{\bf J}}$ with eigenvalues 0.  With this in mind
we denote the basis of states in the subspace of the full Hilbert
space with $j=0$ by $\vert j_1j_2j_3 {\bf 0} \rangle$, where the zero
vector ${\bf 0}$ indicates the vanishing eigenvalues of ${\hat{\bf
J}}$.  These basis states are also eigenstates of the operators
${\hat{\bf J}}_{ij}^2$, for example, ${\hat{\bf J}}_{12}^2 \vert
j_1j_2j_3 {\bf 0}\rangle = j_3(j_3+1) \vert j_1j_2j_3 {\bf
0}\rangle$.  

When the phase of the state $\vert j_1j_2j_3 {\bf 0} \rangle$ is
chosen to agree with Wigner's convention for the phases of the
$3j$-symbols, we have
\begin{equation}
  \left(
  \begin{array}{ccc}
    j_1 & j_2 & j_3 \\
    m_1 & m_2 & m_3 
  \end{array}
  \right) = \langle j_1j_2j_3 m_1m_2m_3 \vert j_1 j_2 j_3 {\bf 0}
  \rangle.
\label{3jdef}
\end{equation}
In this manner we have expressed the $3j$-symbol as a matrix element
connecting the eigenstates of two sets of observables, $({\hat I}_1,
{\hat I}_2, {\hat I}_3, {\hat J}_{1z}, {\hat J}_{2z}, {\hat J}_{3z})$
on the left and $({\hat I}_1, {\hat I}_2, {\hat I}_3, {\hat J}_x,
{\hat J}_y, {\hat J}_z)$ on the right.  Since the second set is
noncommuting, we will require a generalization of
(\ref{abmatrixelement}) to compute the semiclassical approximation to
the $3j$-symbols.

\section{Classical mechanics of the Schwinger model}
\label{cmSchwinger}

The classical mechanics of the Schwinger model must be well understood
in order to carry out a semiclassical analysis.  A general reference
on the classical mechanics of integrable systems from the modern point
of view is Cushman and Bates (1997), where harmonic oscillators in
particular are treated.  

\subsection{The $1j$-model}

We start with the $1j$-model, defining two classical oscillators
$H_\mu = (1/2)(x_\mu^2 + p_\mu^2)$, and $H=\sum_\mu H_\mu$, as in the
quantum case.  The classical configuration space is ${\mathbb R}^2$
and the phase space is ${\mathbb R}^4$.  We introduce complex
coordinates on phase space $z_\mu = (x_\mu +i p_\mu)/\sqrt{2}$ and
${\bar z}_\mu = (x_\mu - i p_\mu)/\sqrt{2}$, where we use an overbar
for complex conjugation.  These are the Weyl symbols of the operators
$a_\mu$, $a^\dagger_\mu$.  The complex coordinates $z_\mu$, ${\bar
z}_\mu$ allow us to identify the phase space ${\mathbb R}^4$ with
${\mathbb C}^2$, that is, knowledge of $z_1$ and $z_2$ allows us to
find all four real coordinates $(x_1,x_2,p_1,p_2)$, since the ${\bar
z}$'s are complex conjugates of the $z$'s.  As we shall see,
coordinates $(z_1,z_2)$, arranged as a 2-component column vector,
transform as a spinor under certain $SU(2)$ transformations.
Variables $z_\mu$ and $i{\bar z}_\mu$ are canonically conjugate ($q$'s
and $p$'s respectively), so that the Poisson bracket of two functions
$f$ and $g$ on phase space can be written,
\begin{eqnarray}
  \{f,g\} &= \sum_\mu \left(
  \frac{\partial f}{\partial x_\mu}
  \frac{\partial g}{\partial p_\mu}-
  \frac{\partial f}{\partial p_\mu}
  \frac{\partial g}{\partial x_\mu}\right) \nonumber \\
  &=
  \sum_\mu \left(
  \frac{\partial f}{\partial z_\mu}
  \frac{\partial g}{\partial (i{\bar z}_\mu)}-
  \frac{\partial f}{\partial (i{\bar z}_\mu)}
  \frac{\partial g}{\partial z_\mu}\right).
  \label{PB}
\end{eqnarray}
   
The basic building blocks of the classical Schwinger model are the function  
\begin{equation}
  I=\frac{1}{2} \sum_\mu {\bar z}_\mu z_\mu = 
  \frac{1}{2}\sum_\mu |z_\mu|^2,
  \label{1jIdef}
\end{equation}
and the three functions
\begin{equation}
J_i = \frac{1}{2} \sum_{\mu\nu} {\bar z}_\mu \, \sigma^i_{\mu\nu} \,
z_\nu,
\label{classJidef}
\end{equation}
for $i=1,2,3$, which define a classical angular momentum vector.  We
also define ${\bf J}^2 = \sum_i J_i^2$.  These functions satisfy the
identity ${\bf J}^2 = I^2$ and the Poisson bracket relations
$\{I,J_i\}=0$, $\{J_i, J_j\} = \sum_k \epsilon_{ijk} J_k$, $\{J_i,
{\bf J}^2\}=0$.

There are two groups of interest that act on the phase space ${\mathbb
R}^4$ or ${\mathbb C}^2$. The first is $U(1)$, generated by $I$.
Hamilton's equations for $I$ are
\begin{equation}
  \frac{d z_\mu}{d \psi} = \frac{\partial I}{\partial (i{\bar z}_\mu)} =
  -\frac{i}{2} z_\mu,
  \qquad
  \frac{d (i{\bar z}_\mu)}{d\psi} = 
  -\frac{\partial I}{\partial z_\mu} = -\frac{1}{2} {\bar z}_\mu,
  \label{Iflowodes}
\end{equation}
where $\psi$ is the parameter of the orbits.  These have the solution
\begin{equation}
  z_\mu(\psi) = \exp(-i\psi/2) \,z_\mu(0), \qquad
  {\bar z}_\mu(\psi) = \exp(i\psi/2) \,{\bar z}_\mu(0).
  \label{Iflow}
\end{equation}
Under the $I$-flow, the two-component spinor $(z_1,z_2)$ just gets
multiplied by an overall phase $\exp(-i\psi/2)$.  Except for
the special initial condition $(z_1,z_2)=(0,0)$ (the origin of phase
space ${\mathbb R}^4$ or ${\mathbb C}^2$), the orbits are circles with
period $4\pi$ with respect to the variable $\psi$.  Henceforth when
citing equations such as (\ref{Iflowodes}) or (\ref{Iflow}) we shall
omit the second half, when it is simply the complex conjugate of the
first half.

We denote a value of $I$ by $j\ge0$.  This is convenient notation, but
in this classical context $j$ is a continuous variable not to be
identified with the quantum number of any operator (see
Sec.~\ref{BSquant}).  Except for the origin $j=0$, the level set $I=j$
(or equivalently, ${\bf J}^2 = j^2$) is the sphere $S^3$, which is
foliated into circles by the action (\ref{Iflow}). This foliation is
precisely the Hopf fibration (Frankel 1997, Nakahara 2003), yielding
the quotient space $S^2 = S^3/S^1$.

The second group acting on phase space is $SU(2)$, whose action is
generated by the $J_i$.  Explicitly, if ${\bf n}$ is a unit vector and
$\theta$ an angle, then the solutions of Hamilton's equations
\begin{equation}
  \frac{d z_\mu}{d\theta} = 
  \frac{\partial ({\bf n}\cdot {\bf J})}{\partial (i{\bar z}_\mu)}=
	-\frac{i}{2} \sum_\nu ({\bf n}\cdot\bsigma)_{\mu\nu} \,
	z_\nu
  \label{ndotJeqns}
\end{equation}
and its complex conjugate are
\begin{equation}
    z_\mu(\theta) = \sum_\nu u({\bf n},\theta)_{\mu\nu}\, z_\nu(0)
  \label{ndotJflow}
\end{equation}
and its complex conjugate.  These are the obvious classical analogs of
equations (\ref{rotatea}); notice that the period in $\theta$ is
$4\pi$.  It is because of this $SU(2)$ action that we say that
coordinates $(z_1,z_2)$ form a spinor.  This classical action of
$SU(2)$ can be understood as a subgroup of the classical group of
linear canonical transformations, $Sp(4)$ (Littlejohn 1986); in
general, $Sp(2N)$ possesses a subgroup $Sp(2N) \cap O(2N)$ that is
isomorphic to $U(N)$, which contains the subgroup $SU(N)$ (in this
case, $N=2$).  When the symplectic matrices lying in the $SU(2)$
subgroup are expressed in the complex basis $(z_\mu, i{\bar z}_\mu)$,
they block diagonalize with $u$ multiplying the $z$'s and ${\bar u}$
multiplying the ${\bar z}$'s.

Equation~(\ref{classJidef}) defines a map (a projection) $\pi:{\mathbb
R}^4 \;(\hbox{\rm or}\; {\mathbb C}^2) \to {\mathbb R}^3$, where
${\mathbb R}^3$ is ``angular momentum space,'' the space with
coordinates $(J_1,J_2,J_3)$.  Here and below we use $\pi$ to denote
this map or its generalization to the $Nj$-model.  The map $\pi$ maps
a larger space onto a smaller one, and so is not one-to-one.  The
inverse image of a point ${\bf J}$ of angular momentum space is a set
of spinors that differ by an overall phase.  It is easy to see that
the definition (\ref{classJidef}) does not depend on the overall phase
of the spinor.  Thus, the inverse image is a circle, except in the
case ${\bf J}=0$ when it is a single point (the origin of
phase space ${\mathbb C}^2$ or ${\mathbb R}^4$).

These circles are precisely the orbits of the $I$-flow (\ref{Iflow}).
Any function $f$ that is constant on these circles projects onto a
well defined function on angular momentum space.  But such functions
are those that Poisson commute with $I$, $\{f,I\}$=0.  This includes
$I$ itself as well as the three $J_i$.  We can write such a function
as $f(z_1,z_2,{\bar z}_1,{\bar z}_2)$ or $f({\bf J})$.  Now if $f$ and
$g$ are any two such functions, then so is their Poisson bracket
$\{f,g\}$, as follows from the Jacobi identity, $\{\{f,g\},I\} =
\{f,\{g,I\}\} + \{g,\{I,f\}\}=0$.  Thus, this Poisson bracket can be
computed directly in angular momentum space without going back to the
bracket (\ref{PB}); the result is the Lie-Poisson bracket,
\begin{equation}
  \{f,g\} = {\bf J} \cdot \left(
  \frac{\partial f}{\partial {\bf J}} \times
  \frac{\partial g}{\partial {\bf J}} \right).
  \label{LiePoisson}
\end{equation}

Interpretations of these spaces may be given in terms of the theory of
``reduction'' (Marsden and Ratiu, 1999).  Angular momentum space is
the Poisson manifold that results from Poisson reduction of the phase
space ${\mathbb R}^4$ under the $U(1)$ action (\ref{Iflow}) generated
by $I$.  It is not by itself an ordinary phase space (symplectic
manifold), which would have an even dimensionality, but it is foliated
into symplectic submanifolds (the symplectic leaves).  In this case
the symplectic leaves are the 2-spheres in angular momentum space,
that is, the level sets ${\bf J}^2 = j^2$, the images under $\pi$ of
the 3-spheres $I=j$ in ${\mathbb R}^4$ or ${\mathbb C}^2$.  Canonical
coordinates on a given 2-sphere ${\bf J}^2 = j^2$ are $(\phi,J_z)$, a
$(q,p)$ pair, where $J_z=j\cos\theta$ and where $(\theta,\phi)$ are
the usual spherical angles in angular momentum space.  Thus we have
\begin{equation}
  dq\wedge dp = d\phi \wedge d(j\cos\theta) = j\,\sin\theta\,d\theta
  \wedge d\phi = j\,d\Omega,
  \label{S2symplform}
\end{equation}
and the symplectic form on a given sphere is $j\,d\Omega$, where
$d\Omega$ is the element of solid angle. This is not the geometrical
solid angle in a Euclidean geometry on angular momentum space, which
would be $j^2 \, d\Omega$.  Another interpretation of angular momentum
space is that it is the dual of the Lie algebra of $SU(2)$, while
$\pi$, given by (\ref{classJidef}), is the momentum map of the
$SU(2)$ action (\ref{ndotJflow}).

We now have three spaces, the ``large phase space'' ${\mathbb R}^4$ or
${\mathbb C}^2$, its image under $\pi$, ``angular momentum space''
${\mathbb R}^3$, and its symplectic leaves, the ``small phase
spaces,'' the 2-spheres ${\bf J}^2 = j^2$. Angular momentum space is
useful for visualizing classical angular momentum vectors, but by
considering inverse projections under $\pi$ the corresponding
geometrical objects in the large phase space can be constructed.
Angular momentum space has been used since the time of the old quantum
theory for visualizing the classical limit of quantum angular momentum
operators; for example, one spoke of an angular momentum vector
``precessing'' around the $z$-direction.  In reality, the
``precession'' defines a manifold of classical states in the small
phase space that is a level set of a complete set of commuting
observables, that is, it is an invariant torus of an integrable system
(just a circle in the $1j$-model, where the commuting observables are
$I$ and $J_z$).

\subsection{The $Nj$-model}

We now consider the classical mechanics of the $Nj$-model, which is
mostly a simple generalization of the $1j$-model. We have $2N$
classical oscillators $H_{r\mu} = (1/2)(x_{r\mu}^2 + p_{r\mu}^2)$; the
configuration space is $({\mathbb R}^2)^N ={\mathbb R}^{2N}$ and the
``large'' phase space is $({\mathbb R}^4)^N={\mathbb R}^{4N}$ or
$({\mathbb C}^2)^N={\mathbb C}^{2N}$.  We define $z_{r\mu}=(x_{r\mu} +
i p_{r\mu})/\sqrt{2}$, ${\bar z}_{r\mu} = (x_{r\mu} -i
p_{r\mu})/\sqrt{2}$, so a point in phase space can be thought of as a
collection of $N$ 2-spinors, $(z_{r1},z_{r2})$, $r=1,\ldots,N$.  We
make the obvious definitions (classical versions of equations
(\ref{Irdef})--(\ref{Jtotalsquareddef})),
\begin{eqnarray}
 I_r &=\frac{1}{2}\sum_\mu |z_{r\mu}|^2,
 \label{classIrdef} \\
 J_{ri} &= \frac{1}{2}\sum_{\mu\nu} {\bar z}_{r\mu}
 \, \sigma^i_{\mu\nu} \, z_{r\nu},
 \label{classJridef}
\end{eqnarray}
as well as ${\bf J}_r^2 = \sum_i J_{ri}^2$, $J_i = \sum_r J_{ri}$ or
${\bf J} = \sum_r {\bf J}_r$, and ${\bf J}^2 = \sum_i J_i^2$.

We denote a value of the functions $I_r$ by $j_r \ge 0$; for positive
values $j_r>0$, $r=1,\ldots,N$, the level set $I_r = j_r$ (or ${\bf
J}_r^2 = j_r^2$) in the phase space $({\mathbb R}^4)^N$ is $S^3 \times
\ldots \times S^3= (S^3)^N$.  The flow generated by $I_r$ for a
specific value of $r$ is just multiplication of the $r$-th spinor
$(z_{r1},z_{r2})$ by a phase factor $\exp(-i\psi_r/2)$, as in
(\ref{Iflow}); the other spinors are not affected.  Thus the $N$
commuting flows generated by all the $I_r$'s constitute a $U(1)^N =
T^N$ action on the large phase space ($T^N$ is the $N$-torus).

Equation~(\ref{classJridef}) defines the projection map $\pi:({\mathbb
C}^2)^N \to ({\mathbb R}^3)^N$, the latter space being ``angular
momentum space'' for the $Nj$-model, with one copy of ${\mathbb R}^3$
for each classical angular momentum vector ${\bf J}_r$.  In view of
its importance, we write out the components of this map explicitly: 
\begin{eqnarray}
  J_{rx} &= \frac{1}{2}({\bar z}_{r1} z_{r2} + {\bar z}_{r2} z_{r1})
  = {\mathop {\rm Re}}({\bar z}_{r1} z_{r2}), 
  \label{Jxeqn} \\
  J_{ry} &= -\frac{i}{2}({\bar z}_{r1} z_{r2} - {\bar z}_{r2} z_{r1})
  ={\mathop {\rm Im}}({\bar z}_{r1} z_{r2}),
  \label{Jyeqn} \\
  J_{rz} &= \frac{1}{2}(|z_{r1}|^2 - |z_{r2}|^2).
  \label{Jzeqn}
\end{eqnarray}
Points of angular momentum space can be visualized as $N$ classical angular
momentum vectors, each living in its own angular momentum space, or $N$
such vectors all in the same 3-dimensional angular momentum space. The
inverse image under $\pi$ of a set of $N$ nonvanishing classical
angular momentum vectors is an $N$-torus in the large phase space,
generated by taking any point in the inverse image (a collection of
$N$ 2-spinors), and multiplying them by $N$ independent, overall phase
factors.  We denote the angles on this torus by $\psi_r$,
$r=1,\ldots,N$, which are the evolution parameters corresponding to
the $I_r$, as in (\ref{Iflow}); thus their periods are $4\pi$.  As
in the $1j$-model, angular momentum space $({\mathbb
R}^3)^N$ is a Poisson manifold, now with Poisson bracket
\begin{equation}
  \{f,g\} = \sum_r {\bf J}_r \cdot \left(
  \frac{\partial f}{\partial {\bf J}_r} \times
  \frac{\partial g}{\partial {\bf J}_r} \right).
  \label{NjLiePoisson}
\end{equation}
The symplectic leaves (the ``small phase spaces'') are the spaces $S^2
\times \ldots \times S^2 = (S^2)^N$ obtained by fixing the values of
$j_1, \ldots, j_N$, with canonical coordinates $(\phi_r,J_{rz})$ on
each sphere.

In the classical $Nj$-model any partial or total sum of the angular
momenta ${\bf J}_r$ generates an $SU(2)$ action on the large phase
space, generalizing equations~(\ref{ndotJeqns}) and (\ref{ndotJflow}) in
the $1j$-model, in that the $SU(2)$ matrix $u$ is applied to all
spinors $(z_{r1},z_{r2})$ whose $r$ values lie in the sum.  For
example, the total ${\bf J}$ rotates all spinors.

These $SU(2)$ actions on the large phase space project to $SO(3)$ actions
on angular momentum space.  Consider, for example, the $SU(2)$ action
generated by the total ${\bf J}$.  Along an orbit in the large phase
space generated by ${\bf n}\cdot {\bf J}$, parameterized by $\theta$,
we can follow the value of ${\bf J}_r$, giving us ${\bf J}_r(\theta)$,
an orbit in the small phase space (the projection under $\pi$ of the
first orbit).  The latter orbit is
\begin{equation}
  J_{ri}(\theta) = \sum_j R({\bf n},\theta)_{ij} \, J_{rj}(0),
  \label{Jrrotation}
\end{equation}
where $R({\bf n},\theta)$ is the $3 \times 3$ rotation associated with
$u({\bf n},\theta)$ according to (\ref{Ruproj}).  This is the
classical analog of (\ref{rotateJr}).  It follows
from (\ref{classJridef}) and the spinor adjoint equation,
$u^\dagger \sigma^i u = \sum_j R_{ij} \, \sigma^j$,
itself equivalent to (\ref{Ruproj}).  Thus, under the $SU(2)$
action on the large phase space generated by ${\bf J}$, the individual
vectors ${\bf J}_r$ rotate in their individual angular momentum spaces
by the corresponding $3 \times 3$ rotation.  For example, $J_z$
rotates all vectors ${\bf J}_r$ about the $z$-axis.  Because of the
two-to-one relation between $SU(2)$ and $SO(3)$, when the orbit in the
large phase space goes around once ($\theta$ goes from 0 to $4\pi$),
the angular momentum vectors go around twice in their individual
angular momentum spaces.  

We may visualize this action as in Fig.~\ref{SU2action}, where $A$
represents a point of angular momentum space (a set of $N$ classical
angular momentum vectors ${\bf J}_r$ in the $Nj$-model).  To obtain
the generic case we assume these vectors are linearly independent (in
particular, none of them vanishes).  In the figure, $T$ is the inverse
image of $A$ under $\pi$, an $N$-torus.  Point $a$ is any specific
point in the large phase space on this $N$-torus, to which the $SU(2)$
rotation $u({\bf n},\theta)$ is applied for $0\le\theta < 4\pi$.  That
is, we treat $a$ as initial conditions for the Hamiltonian flow
generated by ${\bf n}\cdot {\bf J}$, with $\theta$ as the parameter.
This generates the circle $C$ in the large phase space, which amounts
to rotating all $N$ spinors by the same $u({\bf n},\theta)$.  The
projection of the circle $C$ is a circle $D$ in angular momentum space
$({\mathbb R}^3)^N$, that is, all classical vectors ${\bf J}_r$ rotate
about ${\bf n}$ by angle $\theta$.  However, when the circle $C$ is
covered once, circle $D$ is covered twice.  This is because when
$\theta=2\pi$, the spinor rotation $u({\bf n},\theta)=-1$, so all
spinors in the large phase space are just multiplied by $-1$.  This is
illustrated as point $a'$ in the figure, where all spinors are $-1$
times their values at $a$.  Since $-1$ is just a phase factor, both
$a$ and $a'$ project onto the same point $A$ in angular momentum
space.  These are the only two points on $C$ that project onto $A$;
for $\theta$ not a multiple of $2\pi$, the spinor rotation $u({\bf
n},\theta)$ is not a multiplication by a phase factor.

Alternatively, we may apply the entire group $SU(2)$ to the original
point $a$ (not just rotations along a fixed axis).  Then the manifold
$C$ is the orbit of the $SU(2)$ action which is diffeomorphic to
$SU(2)$. The point $a'$ is the image of $a$ under $u=-1$, a specific
element in $SU(2)$, and once again it projects onto the original point
$A$ in the small phase space.  The manifold $D$ is the orbit of point
$A$ under the group $SO(3)$.  In the $1j$-model, it is just a sphere
in angular momentum space (all vectors that can be reached from the
original one by applying all rotations), while in the $Nj$-model for
$N>1$ $D$ is generically diffeomorphic to $SO(3)$ (it is the set of
all classical configurations of $N$ angular momentum vectors that can
be obtained from the original one by applying rigid rotations).

\section{The invariant $jm$-tori}
\label{jmtori}

In this section we continue with the classical point of view,
examining the classical manifolds corresponding to the left side of
the matrix element (\ref{3jdef}).  For this exercise and the rest of
the paper we adopt a $3j$-model ($N=3$).  The manifolds in question
are the level sets of the commuting functions $I_r$, $J_{rz}$,
$r=1,2,3$, or, equally well, of the functions $I_{r\mu} = (1/2)
|z_{r\mu}|^2$ for $r=1,2,3$, $\mu=1,2$, since $I_r = I_{r1}+I_{r2}$
and $J_{rz} = I_{r1} - I_{r2}$.  We denote the level sets by $I_r =
j_r$, $J_{rz} = m_r$ for contour values $j_r$, $m_r$, $r=1,2,3$, or,
equivalently, by
\begin{equation}
	I_{r1} = \frac{1}{2}(j_r+m_r), \qquad
	I_{r2} = \frac{1}{2}(j_r-m_r).
	\label{ia1ia2}
\end{equation}
In spite of the notation, $j_r$ and $m_r$ take on continuous values
and are not necessarily the eigenvalues of any quantum operators.
Since the $I_{r\mu}$ are all nonnegative, we have
\begin{equation}
  j_r \ge 0, \qquad -j_r \le m_r \le j_r,
  \label{jmranges}
\end{equation}
the classical analogs of the usual inequalities in quantum mechanics.

Since each of the six $I_{r\mu}$ is a harmonic oscillator (times
$1/2$), the level set of the $I_{r\mu}$'s is an invariant torus of a
collection of harmonic oscillators.  Generically (for nonzero
amplitude in each oscillator, that is, when none of the quantities
$j_r \pm m_r$ vanishes) this is a 6-torus, upon which the coordinates
may be taken to be the six angles $\theta_{r\mu}$, the variables of
evolution of the $I_{r\mu}$.  The Hamiltonian flow generated by
$I_{r\mu}$ for a specific value of $r$ and $\mu$ just multiplies
$z_{r\mu}$ for the same values of $r$ and $\mu$ by
$\exp(-i\theta_{r\mu}/2)$, while leaving all other $z$'s unaffected.
This is not an overall spinor rotation since the other half of the
spinor containing the given $z_{r\mu}$ is not affected.  If viewed in
the Cartesian $x_{r\mu}$-$p_{r\mu}$ phase plane, this flow is a
clockwise rotation by angle $\theta_{r\mu}/2$, as illustrated in
Fig.~\ref{xpplane}.  The period of the angles $\theta_{r\mu}$ is
$4\pi$.  We agree to measure the angles $\theta_{r\mu}$ from the
positive $x_{r\mu}$ axis, as in the figure, where $z_{r\mu}$ is real
and positive (or zero); this is a specific convention for a set of
canonical coordinates $(\theta_{r\mu}, I_{r\mu})$, $r=1,2,3$,
$\mu=1,2$ on the large phase space.  The volume of the 6-torus with
respect to the measure $d\theta_{11} \wedge \ldots
\wedge d\theta_{32}$ is $(4\pi)^6$.

These tori are also the orbits of the flows generated by the
observables $I_r$, $J_{rz}$.  We denote the evolution variables of the
$I_r$ and the $J_{rz}$ by $\psi_r$ (as above) and $\phi_r$,
respectively.  Each $J_{rz}$ generates an $SU(2)$ rotation about the
$z$-axis on the spinor with the given value of $r$; thus each of the
six angles $(\psi_r,\phi_r)$ has period $4\pi$.  However, when we
allow all six angles $(\psi_r,\phi_r)$ to range from 0 to $4\pi$, the
torus is actually covered eight times.  This can be seen from
Fig.~\ref{SU2action}: a rotation by $2\pi$ in one of the $\phi$'s and
one of the $\psi$'s returns us to the initial point (the path is $a$
to $a'$ along $C$ in Fig.~\ref{SU2action}, then $a'$ to $a$ along
$T$.)  Alternatively, we may consider the canonical transformation
$(\theta_{r\mu}; I_{r\mu}) \to (\psi_r, \phi_r; I_r, J_{rz})$,
generated by
\begin{equation}
  F_2(\theta_{r1}, \theta_{r2}, I_r, J_{rz}) =
  \frac{1}{2}[\theta_{r1}(I_r + J_{rz}) + 
    \theta_{r2}(I_r - J_{rz})],
  \label{F2gf}
\end{equation}
which generates $I_r=I_{r1}+I_{r2}$, $J_{rz}=I_{r1}-I_{r2}$ and 
\begin{equation}
  \psi_r = \frac{1}{2}(\theta_{r1} + \theta_{r2}), \qquad
  \phi_r = \frac{1}{2}(\theta_{r1} - \theta_{r2}),
  \label{psiphiCT}
\end{equation}
so that the Jacobian in the angles is $(1/2)^3=1/8$.  To cover the
torus precisely once we may let the $\psi_r$'s range from $0$ to
$4\pi$ and the $\phi_r$'s from $0$ to $2\pi$, or vice versa; thus the
volume of the torus with respect to $d\psi_1 \wedge d\psi_2 \wedge
d\psi_3 \wedge d\phi_1 \wedge d\phi_2 \wedge d\phi_3$ is
\begin{equation}
	V_{jm} = (2\pi)^3 (4\pi)^3.
	\label{Vjm}
\end{equation} 

The angles $\phi_r$ defined in this way on the large phase space can
be projected onto the small phase space, whereupon they coincide with
the usual azimuthal spherical angles in the individual angular
momentum spaces.  It is clear this must be so to within an additive
constant, since the $J_{rz}$-flow is just an $SO(3)$ rotation about the
$z$-axis in the $r$-th angular momentum space, but by our conventions
even the additive constant comes out right.  To see this we note first
of all that $\phi_r$ for a given $r$ is constant along the $I_s$-flows
for all $s$, since the variables in question are members of a
canonical coordinate system on the large phase space and satisfy
$\{\phi_r, I_s\}=0$.  Thus, $\phi_r$, defined in the large phase
space, projects onto a meaningful function in angular momentum space.
Next, to compute the value of $\phi_r$ for a specific angular momentum
vector ${\bf J}_r$, it suffices to take any point in the 3-torus that
is the inverse image, that is, any value of the angles $\psi_s$ may be
chosen.  For simplicity we take $\psi_r=0$, which implies
$\theta_{r1}=\phi_r$ and $\theta_{r2}=-\phi_r$.  This in turn implies
$z_{r1} = |z_{r1}| \exp(-i\phi_r/2)$, $z_{r2} = |z_{r2}|
\exp(i\phi_r/2)$.  But by equations~(\ref{Jxeqn}) and (\ref{Jyeqn}), these
imply $J_{rx} = J_{r\perp} \cos\phi_r$, $J_{ry} = J_{r\perp}
\sin\phi_r$, where $J_{r\perp} = |z_{r1}||z_{r2}|$.

We shall henceforth call the level set $I_r=j_r$, $J_{rz}=m_r$ the
``$jm$-torus.''  This torus can be projected onto angular momentum
space; we consider the generic case when $j_r \pm m_r \ne 0$ for all
$r$, in which case the $jm$-torus is a 6-torus.  In this case, its
image in angular momentum space is a 3-torus, which, since it is a
surface on which $I_r=j_r$, is also a submanifold of the small phase
space.  This is because the three $\psi_r$ angles just change the
overall phases of the three spinors, without changing their image
under $\pi$, so the three coordinates  on the projected 3-torus are the
angles $\phi_r$.  The 3-torus in angular momentum space can be
visualized as three classical vectors ${\bf J}_r$ in a single angular
momentum space, with specified values of $m_r = J_{rz}$,
``precessing'' about the $z$-axis.  See Fig.~\ref{3Jcones}.  This is
an example of how we shall visualize manifolds in the large phase
space: The $jm$-torus, a six-dimensional manifold in the large phase
space (itself with twelve dimensions), is visualized as three angular
momentum vectors in three dimensional space, as in Fig.~\ref{3Jcones},
defining a 3-torus by varying their azimuthal angles independently,
and each point of this 3-torus is associated with another 3-torus, the
inverse projection under $\pi$ of the given point, which consists of
independently changing the overall phases of the three spinors.  The
6-dimensional $jm$-torus is thus conceived of as the Cartesian product
$T^3 \times T^3$ (it is actually a trivial bundle).

\section{The Wigner manifold}
\label{Wignermanifold}

Now we turn to the right hand side of the matrix element
(\ref{3jdef}), containing the state $\vert j_1j_2j_3 {\bf 0}\rangle$.
This state suggests that we examine the classical manifold upon which
the $I_r$ have definite values, say, $I_r=j_r$, and upon which ${\bf
J}^2=0$.  Again, we do not necessarily identify the $j_r$ with any
quantum numbers, but it is convenient in the following to assume that
none of the $j_r$'s vanishes.  

\subsection{Properties of the Wigner manifold}

Usually the dimensionality of a manifold can be guessed by counting
the constraints that define it, for example, we expect the manifold in
the large (twelve-dimensional) phase space upon which $I_r=j_r$, ${\bf
J}_{12}^2 = j_{12}^2$, $J_z = m$ and ${\bf J}^2 = j^2$, for given
contour values, to be six-dimensional (six constraints on twelve
variables).  Indeed, for most values of $j$ this is correct, and the
manifold in question is a 6-torus (by the Liouville-Arnold theorem,
for certain ranges of the contour values).  These are the invariant
tori that would be involved in the semiclassical treatment of the
addition of three angular momenta, producing a nonzero result (the
case $j\ne0$).  But this naive dimension count only works when the
differentials of the functions in question are linearly independent
(in particular, nonvanishing) on the manifold.  This condition breaks
down when ${\bf J}^2 = j^2 = 0$, since ${\bf J}^2=0$ and ${\bf J}=0$
imply one another, and $d({\bf J}^2)= 2{\bf J}\cdot d{\bf J}=0$.  In
fact, just the four conditions $I_r=j_r>0$, ${\bf J}^2=j^2=0$ define a
six-dimensional manifold in the twelve-dimensional large phase space
(for certain ranges of the $j_r$).  To see this we notice first that
since ${\bf J}^2=0$ implies $J_i=0$, $i=1,2,3$, and $J_z=m=0$ in
particular, the $J_z$ constraint is not independent; neither is the
${\bf J}_{12}^2=j_{12}^2$ constraint, since when ${\bf J}=0$, $j_{12}
= j_3$.  This is just as in the quantum case.  In fact, the manifold
$I_r=j_r$, ${\bf J}^2=0$ is characterized equivalently but better by
$I_r=j_r$, ${\bf J}=0$, since the six differentials $dI_r$ and $dJ_i$
are linearly independent on it.  (Although the $J_i$ vanish on the
manifold in question, their differentials do not.)  Thus, the naive
count of dimensions works with the set $I_r$, $J_i$.

We shall call the manifold $I_r=j_r$, ${\bf J}=0$ in the
large phase space the ``Wigner manifold,'' because
it corresponds to the rotationally invariant state $\vert
j_1j_2j_3 {\bf 0} \rangle$ introduced by Wigner in his definition of
the $3j$-symbols.  The dimensionality of this manifold (six, in the
appropriate ranges of the $j_r$'s) is the same as that of the
invariant tori of any integrable system of six degrees of freedom, and
indeed the same as that of the nearby invariant tori in phase space
corresponding to the level sets of the functions $(I_1,I_2,I_3,{\bf
J}_{12}^2,J_z,{\bf J}^2)$ when $j>0$.  The Wigner manifold, however,
is not a torus.  This is not a contradiction of the Liouville-Arnold
theorem, which requires that the classical observables making up the
level set should Poisson commute.  In the case of the Wigner manifold,
we do have $\{I_r,I_s\}=0$ and $\{I_r,J_i\}=0$, but $\{J_i, J_j\} =
\sum_k \epsilon_{ijk}\, J_k$.  The Hamiltonian vector fields
corresponding to the functions $I_r$, $J_i$ are linearly independent
on the Wigner manifold, but the three $J_i$-flows do not commute (two
Hamiltonian flows commute if and only if their Poisson bracket is a
constant).  The Wigner manifold is, however, an orbit of the
collective action of these Hamiltonian flows (any point can be reached
from any other point by following the flows in some order).  These
facts, information about the topology of the Wigner manifold, and the
required ranges on the contour values $j_r$ will be clarified
momentarily.

The Wigner manifold is also a Lagrangian manifold, like the invariant
tori of an integrable system.  This means that the integral of $p\,dx$
along the manifold is locally independent of path, so an action
function $S(x,A)$ can be defined.  This function in turn is the
solution of the simultaneous Hamilton-Jacobi equations for the
observables $(I_1,I_2,I_3,J_x,J_y,J_z)$, call them $A_i$,
$i=1,\ldots,6$ for short, of which the Wigner manifold is the level
set.  

To prove that the Wigner manifold is Lagrangian, we note that the 
differentials $dA_i$ are linearly independent, so the vector fields
$X_i$ are, too, and span the tangent space to the Wigner manifold at
each point.  Evaluating the symplectic form on these vector fields, we
have $\omega(X_i,X_j) = -\{A_i,A_j\}$.  These Poisson brackets all
vanish except for the $\{J_i,J_j\}$; the latter are nonzero at most
points in phase space, but on the Wigner manifold where ${\bf J}=0$,
these also vanish.  Thus the symplectic form restricted to the Wigner
manifold vanishes, the condition that the Wigner manifold be
Lagrangian.
  
To visualize the Wigner manifold we work our way up from angular
momentum space to the large phase space.  First we attempt to
construct three angular momentum vectors of given positive lengths
$j_1,j_2,j_3$ that add up to the zero vector.  This can be done if and
only if the $j_r$ satisfy the triangle inequalities, whereupon the
values of the $j_r$'s (the lengths of the sides) specify a triangle
that is unique to within orientation.  If we choose a standard or
reference orientation for the triangle, then the three desired vectors
are the vectors running along its sides.  Let us therefore assume the
triangle inequalities are satisfied, and let us choose a standard
orientation for the triangle by placing the ${\bf J}_3$ along the
$z$-axis, ${\bf J}_1$ in the $x$-$z$ plane with $J_{1x}>0$, and ${\bf
J}_2$ in the $x$-$z$ plane with $J_{2x}<0$, as illustrated in
Fig.~\ref{Jtriangle}.  Given any two triangles with the same
(positive) sides, there exists a unique rotation that maps one into
the other; this fact and others regarding triangles are discussed in
the context of the 3-body problem by Littlejohn and Reinsch (1995).
In the present context this means that all classical configurations of
three classical angular momentum vectors of fixed lengths that add up
to the zero vector are related to any one such configuration, such as
the one shown in Fig.~\ref{Jtriangle}, by a unique rotation.  Thus the
manifold of such classical configurations in angular momentum space
${\mathbb R}^3 \times {\mathbb R}^3 \times {\mathbb R}^3$ or in the
small phase space $S^2 \times S^2 \times S^2$ is diffeomorphic to
$SO(3)$.

The Wigner manifold in the large phase space is now the inverse
projection under $\pi$ of this $SO(3)$ manifold in angular momentum
space.  Since the inverse image of any given point of angular momentum
space is a 3-torus in the large phase space (obtained by varying the
overall phases of the three spinors), the Wigner manifold is a 3-torus
bundle over $SO(3)$, and is six-dimensional.  The bundle is
nontrivial.

The Wigner manifold may also be visualized with the help of
Fig.~\ref{Wman}, an elaboration of Fig.~\ref{SU2action}.  It is
assumed that the three $j$'s are positive and satisfy the triangle
inequality.  The lower part of this figure refers to angular momentum
space, while the upper part refers to the large phase space.
Projection $\pi$ maps between the two spaces.  Point $A$ in angular
momentum space is a state of three classical angular momenta of the
given lengths $j_r$ whose vector sum is zero (that is, the angular
momenta define a triangle), in a definite orientation.  To be
specific, let us say that $A$ is the configuration shown in
Fig.~\ref{Jtriangle}.  By applying all $SO(3)$ rotations to $A$ we
generate all orientations of the triangle, of which $B$ in the figure
is one.  The lower circle in the figure represents the manifold of
such configurations, diffeomorphic to $SO(3)$.

The inverse image of any point on this manifold under $\pi$ is a
3-torus in the large phase space.  The 3-tori above points $A$ and $B$
are indicated schematically as lines $T_A$ and $T_B$ in the figure.
Let $a$ be some point on $T_A$.  To be specific, if $A$ is the
configuration of angular momentum vectors shown in
Fig.~\ref{Jtriangle} and the stated conditions on the $j_r$ hold, then
none of the three vectors lies on the negative $z$-axis.  This means
that for any point on $T_A$, $|z_{r1}|^2$ is never zero, since by
equations~(\ref{classIrdef}) and (\ref{Jzeqn}), we have 
\begin{equation}
	|z_{r1}|^2 = j_r + J_{rz}, \qquad
  	|z_{r2}|^2 = j_r - J_{rz},
 	 \label{zrsquare}
\end{equation}
for $r=1,2,3$.  Thus by adjusting the overall phases of the three
spinors $(z_{r1},z_{r2})$, we can make $z_{r1}$ real and positive for
all $r=1,2,3$.  Let this be the point $a$ in Fig.~\ref{Wman}. 

It is notationally tempting to write $m_r$ for the value of $J_{rz}$,
but we shall not do this in the context of the Wigner manifold,
instead reserving the symbol $m_r$ for the contour value of $J_{rz}$
on the $jm$-torus.

Now we apply spinor rotations to point $a$, that is, simultaneous
multiplication of all three spinors $(z_{r1}, z_{r2})$ by the same
element of $SU(2)$.  The orbit thereby generated is a manifold
diffeomorphic to $SU(2)$, as indicated in Fig.~\ref{Wman}.  The
projection of this manifold onto angular momentum space is the surface
$SO(3)$ shown in the figure, that is, all orientations of the triangle
are generated.  For example, in the 3-torus $T_B$ over the angular
momentum triangle with orientation $B$, there is a point $b$ that can
be reached from the given point $a$ by some spinor rotation.  The
spinor rotation in question is one of the two that projects onto the
$SO(3)$ rotation that maps $A$ into $B$, according to (\ref{Ruproj}).
The orbit of reference point $a$ under the $SU(2)$ action therefore
passes through the 3-tori over every possible orientation of the
triangle.  In fact, it passes through each 3-torus twice, since the
$SU(2)$ rotation $u=-1$ is just a phase factor.  This is the meaning
of points $a'$ and $b'$ in the figure, which are related to points $a$
and $b$ by multiplying all three spinors by $-1$.

Thus any point on the Wigner manifold can be reached from the
reference point $a$ by applying some $SU(2)$ rotation, and then
adjusting the overall phases of the three spinors $(z_{r1},z_{r2})$.
The first step is equivalent to following along the Hamiltonian flows
in the large phase space generated by the three $J_i$ (this creates the
$SU(2)$ rotation), while the second is equivalent to following the
Hamiltonian flows generated by the three $I_r$ (this changes the
overall phases of the three spinors).  By letting the rotations range
over all of $SU(2)$ and the three angles $\psi_r$ range from $0$ to
$4\pi$, the Wigner manifold is covered twice.  Thus we obtain
coordinates on the Wigner manifold $(\alpha,\beta,\gamma,
\psi_1,\psi_2,\psi_3)$ (the first three of which are Euler angles on
$SU(2)$).

Solving the simulataneous amplitude transport equations for the six
observables $I_r$, $J_i$, $r,i=1,2,3$ requires us to find an invariant
measure on the Wigner manifold, that is, one invariant under all of
the corresponding Hamiltonian flows.  Details are presented in
Sec.~\ref{ampdet}; for now we just guess that this measure is the
Haar measure on the group $SU(2)$ times the obvious measure on the
3-tori generated by the $I_r$, namely,    
\begin{equation}
  \sin\beta \, d\alpha \wedge d\beta \wedge d\gamma \wedge d\psi_1
  \wedge d\psi_2 \wedge d\psi_3,
  \label{Wignermeasure}
\end{equation}
where $(\alpha,\beta,\gamma)$ are Euler angles on $SU(2)$.  The
integral of this measure over the Wigner manifold is
\begin{equation}
  V_W = \frac{1}{2} (16\pi^2) (4\pi)^3 = 2^9 \pi^5,
  \label{Wignervolume}
\end{equation}
where the $1/2$ compensates for the fact that the Wigner manifold is
covered twice when the Euler angles run over $SU(2)$ and each $\psi_r$
runs from 0 to $4\pi$.

\subsection{Angles related to the shape of the triangle}

Figure~\ref{Jtriangle} defines the angles $\eta_r$ as the angles
opposite vectors ${\bf J}_r$.  Under our assumptions, these angles lie
in the range $0 \le \eta_r \le \pi$.  By projecting all three vectors
onto the directions parallel and orthogonal to each of the vectors in
turn, we obtain a series of identities,
\begin{eqnarray}
  j_1 \cos\eta_2 + j_2 \cos \eta_1 + j_3 &=0, 
  \label{jcosident} \\
  j_1 \sin\eta_2 -j_2 \sin\eta_1 &=0,
  \label{jsinident}
\end{eqnarray}
and four more obtained by cycling indices $1,2,3$.  These allow us to
solve for the cosines of the angles $\eta_r$,
\begin{equation}
  \cos\eta_1 = \frac{j_1^2 -j_2^2 -j_3^2}{2j_2j_3},
  \label{cosetadef}
\end{equation}
and cyclic permutations, which in view of the stated ranges on the
angles allows all three angles $\eta_r$ to be uniquely determined as
functions of $(j_1,j_2,j_3)$.  We shall regard angles $\eta_r$ as
convenient substitutions for these definite functions of the lengths
of the angular momentum vectors.

For the stated ranges on the $\eta_r$, the sines of the angles are
nonnegative and are related to the area $\Delta$ of the triangle, as
follows:
\begin{eqnarray}
  \fl\Delta&= \frac{1}{2} |{\bf J}_1 \times {\bf J}_2| =
  \frac{1}{2} j_1j_2 \sin\eta_3 \nonumber \\
  \fl &= \frac{1}{4} \sqrt{(j_1+j_2+j_3) (-j_1+j_2+j_3) (j_1-j_2+j_3)
    (j_1+j_2-j_3)},
  \label{Deltadef}
\end{eqnarray}
and cyclic permutations.  Some authors define $\Delta$ as the final
square root (without the $1/4$).

\section{Intersections of manifolds}
\label{intersections}

The stationary phase points of the matrix element (\ref{3jdef}) are
the intersections of the $jm$-manifold and the Wigner manifold in the
large phase space.  Thus we must use a version of
(\ref{abmatrixelement1}) for the matrix element, rather than
(\ref{abmatrixelement}).  In this section we study the intersections
of the manifolds, continuing with a classical picture.

If the $jm$-torus and the Wigner manifold in the large phase space
have a common point of intersection, then the projections of these two
manifolds onto angular momentum space must have a common point of
intersection.  The converse is also true:  if the projections have a
point in common, then the inverse image of this point under $\pi$, a
3-torus which is the orbit of the three $I_r$-flows, must contain two
points, one of which belongs to the $jm$-torus, and the other to the
Wigner manifold.  But the 3-torus is the orbit of the $I_r$-flows, and
these flows confine one to both the $jm$-torus and the Wigner
manifold.  Therefore the entire 3-torus is common to both the
$jm$-torus and the Wigner manifold.  Therefore to find intersections
of the $jm$-torus and the Wigner manifold, we may first find the
intersections of their projections under $\pi$.

The $jm$-torus projects onto a set of configurations of
three angular momenta with given lengths and fixed values of $m_r =
J_{rz}$, with arbitrary azimuthal angles, while the Wigner manifold
projects onto configurations with the same lengths in which the vector
sum of the angular momenta vanishes, forming a triangle, with
arbitrary orientation.  Therefore to find a common point between these
two sets of classical angular momenta configurations, we can either
adjust the azimuthal angles of the angular momenta with given $m$
values until a triangle is formed (total ${\bf J}=0$), or we can
rotate a triangle from a given, reference orientation until the $m$
values are the desired ones.  We choose the latter procedure.

Our reference orientation of the triangle is shown in
Fig.~\ref{Jtriangle}, which is indicated schematically as the point
$A$ in Fig.~\ref{Wman}.  We must rotate this reference orientation
to obtain some prescribed values of $m_r$.  These values satisfy the
relations (\ref{jmranges}), so in particular $|m_3|\le j_3$.  Thus by
rotating the triangle in the reference orientation about the $y$-axis
by a unique angle $\beta$, $0\le \beta \le \pi$, defined by
\begin{equation}
  m_3=j_3\cos\beta,
  \label{betadef}
\end{equation}
we guarantee that ${\bf J}_3$ has the right projection.  The result of
this rotation is shown in Fig.~\ref{beta}, for a certain negative
value of $m_3$.

Next we rotate the triangle about the axis ${\bf J}_3$ by an angle
$\gamma$, which does not change ${\bf J}_3$ or its projection, but
which rotates ${\bf J_1}$ and ${\bf J}_2$ in a cone, as illustrated in
Fig.~\ref{gamma}.  We wish to choose the angle $\gamma$ so that ${\bf
J}_2$ has the desired projection $m_2$ onto the $z$-axis.  In
Fig.~\ref{gamma}, ${\bf J}_1$ is not shown, but ${\bf J}_2$ rotates
about the ${\bf J}_3$ direction, its tip sweeping out circle $C_3$.
The circle $C_z$ in the figure is swept out by a vector of length
$j_2$ and projection $m_2$ onto the $z$-axis (this vector is not
shown).  Circles $C_3$ and $C_z$ intersect in two points $Q$ and $Q'$
in the figure, which represent two orientations of the triangle that
have the correct values of both $m_3$ and $m_2$.  Now the orientation
of the triangle is fixed, so there is no more freedom to rotate ${\bf
J}_1$.  In this final orientation the value of $J_{1z}$ is
$-J_{2z}-J_{3z} = -m_2-m_3$, since ${\bf J}=0$ for the triangle.
Either this value of $J_{1z}$ equals the value of $m_1$ associated
with the $jm$-torus, or it does not.  If it does not, then there are
no intersections between the $jm$-torus and the Wigner manifold.
This is just the classical expression of the condition that the matrix
element (\ref{3jdef}) vanishes unless $\sum_r m_r=0$.  Henceforth we
assume that the $m_r$ values for the $jm$-torus do satisfy this
condition.

In this case we may solve for the values of $\gamma$ associated with
points $Q$ and $Q'$ in Fig.~\ref{gamma}.  Writing $R({\bf n},\theta)$
for a 3-dimensional rotation by angle $\theta$ about axis ${\bf n}$,
we have applied the rotation
\begin{equation}
  R({\bf j}_3, \gamma) R({\bf y},\beta) = R({\bf y},\beta)
  R({\bf z},\gamma)
  \label{Rprod}
\end{equation}
to the reference orientation in Fig.~\ref{Jtriangle}, where ${\bf
j}_3$ is the unit vector in the direction ${\bf J}_3$ shown in
Fig.~\ref{beta} (after the first rotation).  In the reference
orientation the vectors are
\begin{equation}
  \fl{\bf J}_1 = j_1 \left(
  \begin{array}{c}
    \sin\eta_2 \\
    0 \\
    \cos\eta_2
  \end{array}\right), \qquad
  {\bf J}_2 = j_2 \left(
  \begin{array}{c}
    -\sin\eta_1 \\
    0 \\
    \cos\eta_1
  \end{array} \right), \qquad
  {\bf J_3}=j_3 \left(
  \begin{array}{c}
    0 \\ 0 \\ 1
  \end{array}\right).
  \label{Jreference}
\end{equation}
After applying rotation (\ref{Rprod}) these become
\begin{eqnarray}
  {\bf J}_1 &= j_1 \left(
  \begin{array}{c}
    \cos\beta\cos\gamma\sin\eta_2 + \sin\beta\cos\eta_2 \\
    \sin\gamma \sin\eta_2 \\
    -\sin\beta\cos\gamma\sin\eta_2 + \cos\beta\cos\eta_2
  \end{array}\right),
	\nonumber \\
  {\bf J}_2 &= j_2 \left(
  \begin{array}{c}
    -\cos\beta\cos\gamma\sin\eta_1 + \sin\beta\cos\eta_1 \\
    -\sin\gamma\sin\eta_1 \\
    \sin\beta\cos\gamma\sin\eta_1 + \cos\beta\cos\eta_1
  \end{array}\right),
  \nonumber \\
  {\bf J}_3 &= j_3 \left(
  \begin{array}{c}
    \sin\beta \\
    0 \\
    \cos\beta
  \end{array}\right).
  \label{Jrotated}
\end{eqnarray}
We have already solved for $\beta$ in (\ref{betadef}); we may now
solve for $\gamma$ by demanding either $J_{1z}=m_1$ or $J_{2z}=m_2$.
These lead to
\begin{equation}
  \cos\gamma = 
  \frac{j_1 \cos\beta\cos\eta_2 -m_1}{j_1\sin\beta\sin\eta_2} =
  \frac{m_2-j_2 \cos\beta\cos\eta_1}{j_2\sin\beta\sin\eta_1}.
  \label{gammadef}
\end{equation}
These two conditions are equivalent (under the assumption $\sum_r
m_r=0$), as follows from the identities
(\ref{jcosident})--(\ref{cosetadef}).  If the common value of the two
expressions on the right hand side of (\ref{gammadef}) lies in the
range $(-1,+1)$, then there are two real angles $\gamma$ satisfying
(\ref{gammadef}), corresponding to the two points $Q$ and $Q'$ in
Fig.~\ref{gamma}.  In this case the two manifolds have real
intersections, and we are in the classically allowed region for the
$3j$-symbol.  We let $\gamma$ represent the root (the ``principal
branch'') in the range $[0,\pi]$, and $-\gamma$ the root (the
``secondary branch'') in the range $[-\pi,0]$.  Note that
$\sin\gamma\ge0$ ($\le0$) on the principal (secondary) branch.  If the
right hand side of (\ref{gammadef}) lies outside the range
$[-1,1]$, then there are two complex roots for $\gamma$.  In this case
the two manifolds have no real intersections, but they do have complex
ones.  Only one of the two complex roots is picked up by the contour
of integration used in obtaining the matrix element
(\ref{abmatrixelement}) or (\ref{abmatrixelement1}), resulting in an
exponentially decaying expression for the matrix element.  In this
case we are in the classically forbidden region of the $3j$-symbol.
In the following for simplicity we assume we are in the classically
allowed region.

The points $Q$ and $Q'$ in Fig.~\ref{gamma} represent values of ${\bf
J}_2$ in a single angular momentum space ${\mathbb R}^3$.  Taken with
the values of ${\bf J}_1$ and ${\bf J}_3$, they specify points, call
them $P$ and $P'$, in the combined angular momentum space ${\mathbb
R}^3 \times {\mathbb R}^3 \times {\mathbb R}^3$ that lie on the
intersection of the projections of the $jm$-torus and the Wigner
manifold onto that space.  Then by applying rotations about the
$z$-axis to $P$ and $P'$, we generate a pair of circles in angular
momentum space.  Such rotations change neither the $z$-projections
of the three vectors nor their vector sum (zero).  This is obviously a
reflection of the fact that the operator $J_z$ defining the state on
the right side of (\ref{3jdef}) is a function of the operators
$(J_{1z}, J_{2z}, J_{3z})$ defining the state on the left.  Thus the
projections of the $jm$-manifold and Wigner manifold under $\pi$
intersect generically in a pair of circles. 

Thus the intersection of the $jm$-torus and the Wigner manifold in
the large phase space is the inverse image of this pair of circles
under $\pi$, generically a pair of 3-torus bundles over a circle.
Since the $I_r$-flows and the $J_z$-flow commute, these bundles are
trivial, in fact each is a 4-torus, on which coordinates are
$(\psi_1,\psi_2,\psi_3,\phi)$, where $\phi$ is the angle of evolution
along the $J_z$-flow.  The volume of either one of the 4-tori with
respect to the measure $d\psi_1 \wedge d\psi_2 \wedge d\psi_3 \wedge
d\phi$ is
\begin{equation}\
	V_I = \frac{1}{2} (4\pi)^4,
	\label{VIdef}
\end{equation}
the factor of $1/2$ being explained by Fig.~\ref{SU2action}.  

The $jm$-torus and the Wigner manifold, both six-dimensional,
intersect in a 4-torus because the the lists of functions defining the
two manifolds, $(I_1,I_2,I_3,J_{1z},J_{2z},J_{3z})$ and
$(I_1,I_2,I_3,J_x,J_y,J_z)$, have three functions in common while
$J_z$ in the second list is a function of $(J_{1z},J_{2z},J_{3z})$ in
the first list.  Below we will transform the functions to make both
lists have explicitly four variables in common (see Eq.~(\ref{JzCT})).
  
\section{Action integrals}
\label{actionintegrals}

Action integrals on the $jm$-torus and the Wigner manifold are needed
for the phases in expresssions such as (\ref{abmatrixelement1}).  We
only need the action function at some point on the intersection
between the two manifolds, which gives us a lot of choice since that
intersection is a 4-torus.  We continue with a classical picture in
this section.

\subsection{Choosing reference points} 

Action integrals are defined relative to some initial or reference
point on each manifold.  For the $jm$-torus, a convenient point is the
one where $\theta_{r\mu}=0$, $r=1,2,3$, $\mu=1,2$, that is, the point
where each $z_{r\mu}$ is real and nonnegative, as explained in
Sec.~\ref{jmtori}.  According to (\ref{ia1ia2}), the spinors at
this reference point are given explicitly by
\begin{equation}
	\left(\begin{array}{c}
	z_{r1} \\ z_{r2}
	\end{array}\right) =
	\left(\begin{array}{c}
	\sqrt{j_r + m_r} \\
	\sqrt{j_r - m_r}
	\end{array}\right).
	\label{jmref}
\end{equation}
The projection of this point onto angular momentum space is a set of
vectors ${\bf J}_r$, $r=1,2,3$ of given lengths $j_r$ that lie in the
$x$-$z$ plane, with $J_{rz}=m_r$ and $J_x \ge 0$, as shown by
(\ref{Jxeqn})--(\ref{Jzeqn}).  Such vectors are illustrated in
Fig.~\ref{3Jcones}.

As for the Wigner manifold, it is convenient to take the reference point
to be point $a$ in Fig.~\ref{Wman}, which is discussed in
Sec.~\ref{Wignermanifold}.   This point projects onto the standard
orientation of the triangle, point $A$ in Fig.~\ref{Wman}, where the
angular momentum vectors have the values shown in
(\ref{Jreference}).  At the point $a$, $z_{r1}$ is real and
positive for all $r$, as explained in Sec.~\ref{Wignermanifold}.  For
example, for $r=1$ this assumption combined with (\ref{zrsquare})
implies $z_{11}=\sqrt{j_1+J_{1z}}$, which by (\ref{Jreference})
becomes $z_{11}=\sqrt{j_1(1+\cos\eta_2)}=\sqrt{2j_1}\cos\eta_2/2$.
Then (\ref{Jyeqn}) and $J_{1y}=0$ imply that $z_{12}$ is purely
real, and (\ref{Jxeqn}) allows us to solve for $z_{12}$ in terms
of $J_{1x}$, given by (\ref{Jreference}), producing finally
$z_{12}=\sqrt{2j_1}\sin\eta_2$.  Proceeding similarly with the other
two spinors $r=2,3$, we obtain the three spinors at the reference
point $a$ on the Wigner manifold,
\begin{eqnarray}
	\fl\left(\begin{array}{c}
	z_{11} \\ z_{12}
	\end{array}\right) &=
	\sqrt{2j_1} \left(\begin{array}{c}
	\cos\eta_2/2 \\
	\sin\eta_2/2
	\end{array}\right), \qquad
	\left(\begin{array}{c}
	z_{21} \\ z_{22}
	\end{array}\right) =
	\sqrt{2j_2} \left(\begin{array}{c}
	\cos\eta_1/2 \\
	-\sin\eta_1/2
	\end{array}\right), 
	\nonumber \\
	[3pt]
	\fl\left(\begin{array}{c}
	z_{31} \\ z_{32}
	\end{array}\right) &=
	\sqrt{2j_3} \left(\begin{array}{c}
	1 \\ 0
	\end{array}\right).
	\label{Wref}
\end{eqnarray}

Now to obtain a point common to both the $jm$-torus and the Wigner
manifold, we apply the spinor rotation 
\begin{equation}
  u({\bf y},\beta) u({\bf z},\gamma)= \left(
  \begin{array}{cc}
    e^{-i\gamma/2}\cos\beta/2 & -e^{i\gamma/2}\sin\beta/2 \\
    e^{-i\gamma/2}\sin\beta/2 & e^{i\gamma/2}\cos\beta/2
  \end{array} \right)
  \label{uprod}
\end{equation}
to the reference spinors $(z_{r1},z_{r2})$, $r=1,2,3$,
in (\ref{Wref}), where Euler angles $\beta$
and $\gamma$ are defined by (\ref{betadef}) and
(\ref{gammadef}).  We obtain either the principal branch or the
secondary one by taking $\gamma\ge0$ or $\gamma\le0$, respectively.  
The spinor rotation (\ref{uprod}) induces the $3\times 3$
rotation on the angular momentum vectors shown in (\ref{Rprod}).
Thus we obtain the spinors at the common point of intersection between
the $jm$-torus and the Wigner manifold,
\begin{eqnarray}
  	\fl\left(\begin{array}{c} 
	z_{11} \\ z_{12} \end{array} \right) &=
  	\sqrt{2j_1}  \left( \begin{array}{c}
	e^{-i\gamma/2} \cos\beta/2 \cos\eta_2/2 - 
    	e^{i\gamma/2} \sin\beta/2 \sin\eta_2/2 
	\label{ip1} \\
    	[3pt]
    	e^{-i\gamma/2} \sin\beta/2 \cos\eta_2/2 + 
    	e^{i\gamma/2 } \cos\beta/2 \sin\eta_2/2
  	\end{array}\right), \\
  	\fl\left(\begin{array}{c} 
	z_{21} \\ z_{22} \end{array}\right) &=
  	\sqrt{2j_2}  \left( \begin{array}{c}
    	e^{-i\gamma/2} \cos\beta/2 \cos\eta_1/2 + 
	e^{i\gamma/2} \sin\beta/2 \sin\eta_1/2 
	\label{ip2} \\
	[3pt]
    	e^{-i\gamma/2} \sin\beta/2 \cos\eta_1/2 - 
	e^{i\gamma/2} \cos\beta/2 \sin\eta_1/2
  	\end{array}\right), \\
  	\fl\left(\begin{array}{c} 
	z_{31} \\ z_{32} \end{array}\right) &=
  	e^{-i\gamma/2} \sqrt{2j_3}  \left(
  	\begin{array}{c}
    	\cos\beta/2 \\ \sin\beta/2
  	\end{array}\right).
	\label{ip3}
\end{eqnarray}
One can easily check using (\ref{Jxeqn})--(\ref{Jzeqn}) that
these spinors project onto the angular momentum vectors in
(\ref{Jrotated}). 

\subsection{Computing the actions} 

In computing action integrals we use the identity,
\begin{equation}
	\sum_{r\mu} p_{r\mu} \, dx_{r\mu} =
	\frac{i}{2} \sum_{r\mu} 
	({\bar z}_{r\mu} \, dz_{r\mu} -
	z_{r\mu} \, d{\bar z}_{r\mu}) +
	\frac{1}{2} d \sum_{r\mu} 
	x_{r\mu} p_{r\mu}.
	\label{diffformident}
\end{equation}
The integral of the left hand side is the usual action one would need
for wave functions $\psi(x_{11}, \ldots, x_{32})$, but it can be
replaced by the integral of the first differential form on the right,
for the following reason.  First, the integral of the exact
differential on the right contributes the difference in the function
$(1/2)\sum_{r\mu} x_{r\mu} p_{r\mu}$ between the initial and final
points.  But the final point is the common point of intersection
between the $jm$-torus and the Wigner manifold, so this contribution
cancels when we subtract actions as in (\ref{abmatrixelement1}).  As
for the initial points on the two manifolds, these have been chosen
(see Eqs.~(\ref{jmref}) and (\ref{Wref})) so that all $z_{r\mu}$ are
purely real, or $p_{r\mu}=0$.  Thus the function in question vanishes
at the initial points.  As for the integral of the first term on the
right of (\ref{diffformident}), it can be written
\begin{equation}
	S=
	\mathop{\rm Im} \int \sum_{r\mu}
	z_{r\mu} \, d{\bar z}_{r\mu}.
	\label{zaction}
\end{equation}

For the action on the $jm$-torus between initial point (\ref{jmref})
and final point (\ref{ip1})--(\ref{ip3}), we follow a path consisting
of flows of the functions $I_{r\mu} = (1/2)|z_{r\mu}|^2$ taken one at
a time by angles $\theta_{r\mu}$.  Along the $I_{r\mu}$-flow we have
$d{\bar z}_{r\mu}/ d\theta_{r\mu} = (i/2) {\bar z}_{r\mu}$, so the
contribution to $S$ is
\begin{equation}
	\mathop{\rm Im} \int_0^{\theta_{r\mu}}
	\frac{i}{2} |z_{r\mu}|^2 \, d\theta_{r\mu} =
	I_{r\mu} \theta_{r\mu},
	\label{Irmucontrib}
\end{equation}
since $I_{r\mu}$ is constant along its own flow and since
$\theta_{r\mu}=0$ at the reference point.  Thus the total action
between initial and final points on the $jm$-torus is
\begin{equation}
	S_{jm}=\sum_{r\mu} I_{r\mu} \theta_{r\mu}.
	\label{jmaction}
\end{equation}
Under the canonical transformation (\ref{F2gf}) this becomes
\begin{equation}
	S_{jm} = \sum_r (I_r \psi_r + J_{rz} \phi_r) =
	\sum_r (j_r \psi_r + m_r \phi_r),
	\label{jmaction1}
\end{equation}
where in the final form we replace $I_r$ and $J_{rz}$ by their values
on a given $jm$-torus.  

The angles $\theta_{r\mu}$ or $(\psi_r,\phi_r)$ are the coordinates
of the final point specified by Eqs.~(\ref{ip1})--(\ref{ip3}).  The
solutions of Hamilton's equations for the $I_r$-flow can be written
$z_{r\mu}(\theta_{r\mu}) = z_{r\mu}(0) \exp(-i\theta_{r\mu}/2)$ (see
Fig.~\ref{xpplane}) where the initial conditions are real and
nonnegative, so we have $\theta_{r\mu} = 2 \mathop{\rm arg} {\bar
z}_{r\mu}$.  Combining this and (\ref{psiphiCT}), we can write the
action on the $jm$-torus as 
\begin{equation}
	\fl S_{jm} = 2 \sum_{r\mu} I_{r\mu} 
	\mathop{\rm arg} {\bar z}_{r\mu} =
	\sum_r j_r \mathop{\rm arg}
	({\bar z}_{r1} {\bar z}_{r2}) + 
	\sum_r m_r \mathop{\rm arg}
	({\bar z}_{r1} z_{r2}).
	\label{jmaction2}
\end{equation}
Using Eqs.~(\ref{ip1})--(\ref{ip3}), this can be written,
\begin{eqnarray}
	S_{jm} &= j_3\gamma
	+j_1 \mathop{\rm arg}(
	\cos\beta \sin\eta_2 + 
	\sin\beta \cos\gamma \cos\eta_2 +
	i\sin\beta \sin\gamma) \nonumber \\
	&+j_2 \mathop{\rm arg}(
	-\cos\beta \sin\eta_1 +
	\sin\beta \cos\gamma \cos\eta_1 +
	i\sin\beta \sin\gamma) \nonumber \\
	&+m_1 \mathop{\rm arg}(
	\sin\beta \cos\eta_2 +
	\cos\beta \cos\gamma \sin\eta_2 +
	i\sin\gamma \sin\eta_2) \nonumber \\
	&+m_2 \mathop{\rm arg}(
	\sin\beta \cos\eta_1 -
	\cos\beta \cos\gamma \sin\eta_1 -
	i\sin\gamma\sin\eta_1).
	\label{jmaction3}
\end{eqnarray}

Here we have used the rule $\mathop{\rm arg}(ab) = \mathop{\rm arg}a +
\mathop{\rm arg} b$, which is only valid for certain choices of branch
of the arg function.  A more careful analysis shows that
(\ref{jmaction3}) is the correct action along a certain path from the
initial to final point (the principal branch) on the $jm$-torus if the
range of the arg function is taken to be $[-\pi,\pi)$.  (The path is
defined by $\gamma \le \theta_{11}, \theta_{22} \le 2\pi$, $-\gamma
\le \theta_{12}, \theta_{21} \le \gamma$, that is, one integrates from
0 to these final $\theta$ values.)  In particular, this means that
$\psi_1, \psi_2,\psi_3,\phi_1$ all lie in $[0,\pi]$, while $\phi_2$
lies in $[-\pi,0]$.  These ranges on angles $\phi_1$, $\phi_2$ are
also evident from Fig.~\ref{gamma}.  Similarly, for the secondary
branch ($-\pi \le \gamma \le 0$, $\sin\gamma \le 0$) there exists a
path such that with the same range on the arg function
(\ref{jmaction3}) is still correct.  With these understandings, the
values of $S_{jm}$ on the two branches differ by a sign.  We shall
henceforth write $S_{jm}$ ($-S_{jm}$) for the principal (secondary)
branch.

Equation~(\ref{jmaction}) can also be written in terms of $\cos^{-1}$
functions.  We note that $\mathop{\rm arg}({\bar z}_{11} {\bar
z}_{12}) = \cos^{-1}[ \mathop{\rm Re} ({\bar z}_{11} {\bar z}_{12})
/|z_{11} z_{12}|]$ and that $|z_{11} z_{12}| = (j_1^2 -m_1^2)^{1/2}$,
etc.  We can also use (\ref{gammadef}) to eliminate $\cos\gamma$.  For
the principal branch ($\gamma\ge0$) this gives
\begin{eqnarray}
	\fl S_{jm} &= j_1 \cos^{-1} \left(
	\frac{j_1 \cos\beta - m_1 \cos\eta_2}
	{\sin\eta_2 J_{1\perp}}\right) +
	j_2 \cos^{-1} \left(
	\frac{m_2\cos\eta_1 - j_2 \cos\beta}
	{\sin\eta_1 J_{2\perp}}\right) \nonumber \\
	[3pt]
	\fl &+j_3 \cos^{-1} \left(
	\frac{j_1\cos\beta\cos\eta_2 -m_1}
	{j_1\sin\beta\sin\eta_2}\right) +
	m_1\cos^{-1}\left(
	\frac{j_1\cos\eta_2 -m_1\cos\beta}
	{\sin\beta J_{1\perp}}\right) \nonumber \\
	[3pt]
	\fl &-m_2 \cos^{-1} \left(
	\frac{j_2\cos\eta_1 -m_2 \cos\beta}
	{\sin\beta J_{2\perp}}\right),
	\label{jmaction4}
\end{eqnarray}
where
\begin{equation}
	J_{r\perp} = \sqrt{j_r^2 - m_r^2},
	\label{Jrperpdef}
\end{equation}
and where the range of the $\cos^{-1}$ function is $[0,\pi]$.
Finally, by using Eqs.~(\ref{cosetadef}) and (\ref{Deltadef}) these can
be written explicitly in terms of the parameters $j_r$, $m_r$.  The
result has the form of (\ref{jmaction1}), where
\begin{equation}
	\psi_1 = \cos^{-1} \left(
	\frac{j_1^2(m_3 - m_2) + m_1(j_3^2 - j_2^2)}
	{4\Delta J_{1\perp}}\right),
	\label{psi1def}
\end{equation}
and cyclic permutations of indices, and where
\begin{equation}
	\fl \phi_1 = \cos^{-1}\left(
	\frac{j_2^2 - j_3^2 -j_1^2 - 2m_1m_3}
	{2J_{1\perp} J_{3\perp}}\right), 
	\; \phi_2 = -\cos^{-1}\left(
	\frac{j_1^2 -j_3^2 -j_2^2 -2m_2m_3}
	{2 J_{2\perp} J_{3\perp}}\right),
	\label{phi12def}
\end{equation}
and where $\phi_3=0$.  

Now we consider the action on the Wigner manifold between the initial
point (\ref{Wref}) and the final point (\ref{ip1})--(\ref{ip3}).  The
path between these points is made up of the product of rotations
(\ref{uprod}), so we consider the action integral (\ref{zaction})
along a rotation by angle $\theta$ generated by ${\bf n}\cdot {\bf
J}$.  Hamilton's equations (see Eq.~(\ref{ndotJeqns})) are $d{\bar
z}_{r\mu}/d\theta = (i/2) \sum_\nu {\bar z}_{r\nu} ({\bf n} \cdot
\bsigma)_{\nu\mu}$, so by (\ref{zaction}) we have
\begin{equation}
	\fl S=\mathop{\rm Im} \int_0^\theta 
	\frac{i}{2} \sum_{r\mu\nu} {\bar z}_{r\nu}
	({\bf n} \cdot \bsigma)_{\nu\mu} z_{r\mu} \,
	d\theta =
	\int_0^\theta ({\bf n} \cdot {\bf J})\, 
	d\theta =
	({\bf n} \cdot {\bf J})\theta=0,
	\label{rotaction}
\end{equation}
where we use (\ref{classJridef}), the fact that ${\bf n} \cdot
{\bf J}$ is constant along its own flow, and the fact that ${\bf J}=0$
on the Wigner manifold.  The rotational action vanishes.

Thus the phase of the matrix element (\ref{3jdef}) is determined
entirely by the action integral along the $jm$-torus, that is, to
within a sign it is given by Eqs.~(\ref{jmaction})--(\ref{phi12def}).
This is the phase function determined previously by Ponzano and Regge,
Miller, and others, and we see that it is essentially a simple
combination of the phases of the Schwinger oscillators.  We have, however,
determined this phase function entirely within a classical model, that
is, without imposing any quantization conditions on the manifolds.  

\section{Bohr-Sommerfeld quantization}
\label{BSquant}

We do not need Bohr-Sommerfeld approximations to the eigenvalues of
the operators involved in the $3j$-symbols because those eigenvalues
are known exactly.  We must, however, quantize the $jm$-torus and the
Wigner manifold, to obtain the wave functions whose scalar product is
the $3j$-symbol.  We also need the Bohr-Sommerfeld rules to make the
connection between the contour values for various classical functions
and the standard quantum numbers of the associated operators.

To quantize a Lagrangian manifold we must first find the generators of
the fundamental or first homotopy group of the manifold, that is, a
set of closed contours in terms of which all closed contours can be
generated by concatening curves.  In the following we shall call these
generators ``basis contours,'' although technically the fundamental
group, even when Abelian, is a group and not a vector space.  For
example, in the familiar case of the invariant $n$-tori of integrable
systems of $n$ degrees of freedom, the fundamental group is ${\mathbb
Z}^n$, that is, an arbitrary closed contour is expressed as a ``linear
combination'' of the $n$ basis contours with integer coefficients.
The $n$ basis contours themselves go around the torus once in the $n$
different directions.

After finding the basis contours, we compute the total phase
associated with each of them, the sum of an action, the integral of
$p\,dq$ around the contour, and a Maslov phase, which is $-\pi/2$
times the Maslov index of the loop.  Both these phases are topological
invariants and are additive when loops are concatenated.  Then we
demand that the total phase be a multiple of $2\pi$; this is the
consistency condition on the semiclassical wave function that selects
out certain manifolds as being ``quantized.''

The Lagrangian manifolds we are interested in are level sets of a set
of classical functions that are the principal symbols of a set of
operators, which in our application need not commute.  Quantized
Lagrangian manifolds support wave functions that are approximate
eigenfunctions of the set of operators.  The corresponding eigenvalues
are the contour values of the principal symbols, to within errors of
order $\hbar^2$.

\subsection{Quantizing the $jm$-tori}

In the case of the $jm$-tori, whose fundamental group is ${\mathbb
Z}^6$, we are dealing with the eigenfunctions of a set of independent
harmonic oscillators, so the problem could not be more elementary from
a semiclassical standpoint.  There are, however, interesting issues
that arise.  Let the complete set of commuting quantum observables be
$({\hat I}_r, {\hat J}_{rz})$, $r=1,2,3$, defined in (\ref{Irdef})
and (\ref{Jridef}).  Let us denote the Weyl symbol of an operator
${\hat A}$ by $\mathop{\rm sym}({\hat A})$.  Then we have
\begin{equation}
	\mathop{\rm sym}({\hat I}_r) =
	    I_r -\frac{1}{2}, \qquad
	\mathop{\rm sym}({\hat J}_{rz}) =
	J_{rz},
	\label{IJsymbols}
\end{equation}
where $I_r$ and $J_{rz}$ (the classical functions) are defined by
Eqs.~(\ref{classIrdef}) and (\ref{classJridef}).  The operator ${\hat
I}_r$ violates our assumption in Sec.~\ref{scwfis} that the commuting
operators defining our integrable system should have Weyl symbols that
are even power series in $\hbar$ (since the $-1/2$ is of order
$\hbar$).  Our assumption is valid for the harmonic oscillators ${\hat
H}_r = (1/2)\sum_\mu ({\hat x}_{r\mu}^2 + {\hat p}_{r\mu}^2)$, but in
defining ${\hat I}_r$ in (\ref{Irdef}) we have subtracted the zero
point energy, a constant of order $\hbar$, ${\hat I}_r = (1/2)({\hat
H}_r - 1)$, so that the eigenvalues of ${\hat I}_r$ would be the
conventional quantum numbers $j_r$ for an angular momentum, and so
that the identity (\ref{JIidentity}) would have a familiar form.  In
the following we shall take the principal symbol of ${\hat I}_r$ to be
the whole symbol, including the $-1/2$.  This achieves the same
results we would have had if we had worked with ${\hat H}_r$ instead
of ${\hat I}_r$ and defined the principal symbol as the leading term
in $\hbar$, as in Sec.~\ref{scwfis}, since $\mathop{\rm sym}({\hat
H}_r) = 2I_r$.  We must be careful, however, since the principal
symbol of ${\hat I}_r$ is not $I_r$.  For most of the other operators
we shall use, the principal symbol is obtained simply by removing the
hat (for example, $J_{rz}$ above).

The basis contours on the $jm$-torus are most easily expressed as the
contours on which each of the angles $\theta_{r\mu}$ is allowed to go
from 0 to $4\pi$ while all other $\theta_{r\mu}$'s are held fixed.  We
may also use any linear combination of these contours with integer
coefficients and unit determinant.  In terms of the angles $\psi_r$,
$\phi_r$, given by (\ref{psiphiCT}), a convenient choice is to
take one basis contour as the path on which one $\psi_r$ goes from $0$
to $4\pi$ while all others $\psi_r$'s and all $\phi_r$'s are held
fixed; this is following the $I_r$-flow for elapsed angle $\psi_r=4\pi$.
A second basis contour may be taken to be the path on which $\psi_r$
goes from 0 to $2\pi$, and then $\phi_r$ goes from 0 to $2\pi$; the
two legs involve following the $I_r$-flow and then the $J_{rz}$-flow,
each for elapsed angle $2\pi$.  Doing this for $r=1,2,3$ gives us 
six basis contours on the $jm$-torus.

The action along the first basis contour is computed as in
(\ref{Irmucontrib}).  Hamilton's equations for $I_r$ are $d{\bar
z}_{r\mu}/d\psi_r = (i/2){\bar z}_{r\mu}$, so we obtain $S_{r1}=4\pi
I_r$, where $S_{r1}$ refers to the action along the first basis
contour.  For the second basis contour, the first leg contributes an
action $2\pi I_r$, while the second leg, which follows the flow
generated by $J_{rz}$, is a rotation whose action may be computed as
in (\ref{rotaction}), but with ${\bf J}$ replaced by ${\bf J}_r$ since
we do not sum over $r$.  Thus the final answer does not vanish (${\bf
J}_r$ is nonzero on the $jm$-torus), and the contribution from the
second leg is $2\pi J_{rz}$, or $S_{r2} = 2\pi(I_r + J_{rz})$.
Altogether, we have
\begin{equation}
	S_{r1} = 4\pi j_r, \qquad
	S_{r2} = 2\pi(j_r + m_r),
	\label{jmactions}
\end{equation}
where the 1 and 2 refer to the first and second basis contours
associated with a particular value of $r$, and where we have replaced
$I_r$ and $J_{rz}$ by their contour values $j_r$ and $m_r$ on the
$jm$-torus.

Next we need the Maslov indices along the two basis contours.  Here we
follow the computational method described in Littlejohn and Robbins
(1987), which uses the determinant of complex matrices and which is
based ultimately on Arnold (1967).  Similar techniques are discussed
by Mishchenko \etal\ (1990). The method works for finding Maslov
indices along closed curves on orientable Lagrangian manifolds in
${\mathbb R}^{2n}$.  To describe the method we adopt a general
notation, in which global coordinates on phase space are $(q_1,
\ldots, q_n,p_1, \ldots, p_n)$.  We suppose that there exists a set of
$n$ vector fields on the Lagrangian manifold, linearly independent at
each point, so that they span the Lagrangian tangent
plane at each point.  In our applications, these are the Hamiltonian
vector fields associated with a set of functions $(A_1, \ldots, A_n)$.
We consider the rate of change of the quantities $q_i -i p_i$ along
the $j$-th vector field, which is the Poisson bracket
$\{q_i-ip_i,A_j\}$.  The set of these Poisson brackets forms an $n
\times n$ complex matrix $M_{ij}$ that is never singular, so $\det M$
traces out a closed loop in the complex plane without passing through
the origin when we go around a closed loop on the Lagrangian manifold.
Then the Maslov index $\mu$ associated with this loop is given by
\begin{equation}
	\mu = 2 \mathop{\rm wn} \det M,
	\label{mudef}
\end{equation}
where wn refers to the winding number of the loop in the complex
plane, reckoned as positive in the counterclockwise direction.  The
winding number is invariant when $M_{ij}$ is multiplied by any nonzero
complex constant (or constant matrix), so such constants can be
dropped in the calculation.

For the $jm$-torus, we identify the $q$'s and $p$'s with the
coordinates $x_{r\mu}$ and $p_{r\mu}$, and the $A$'s with the
functions $(I_1,I_2,I_3, J_{1z},J_{2z},J_{3z})$.  Then we can replace
$q_i-ip_i$ by ${\bar z}_{r\mu}$, dropping the $1/\sqrt{2}$.  The
needed matrix elements are
\begin{equation}
	\eqalign{
	\{{\bar z}_{r\mu}, I_s\} &= 
	i\frac{\partial I_s}{\partial z_{r\mu}} =
	\frac{i}{2} \delta_{rs} \, {\bar z}_{r\mu}, \\
	\{{\bar z}_{r1}, J_{sz}\} &=
	i\frac{\partial J_{sz}}{\partial z_{r1}} =
	\frac{i}{2} \delta_{rs} \, {\bar z}_{r1}, \\
	\{{\bar z}_{r2}, J_{sz}\} &=
	i\frac{\partial J_{sz}}{\partial z_{r2}} =
	-\frac{i}{2} \delta_{rs} \, {\bar z}_{r2}.}
	\label{Mijjm}
\end{equation}
We drop the constant $i/2$ on the right hand side, and choose the
ordering $(I_1,J_{1z}, I_2,J_{2z}, I_3,J_{3z})$ for the functions.
Then the $6\times 6$ matrix block diagonalizes into three $2 \times 2$
blocks, and we find
\begin{equation}
	\det M = {\bar z}_{11} {\bar z}_{12} {\bar z}_{21}
	{\bar z}_{22} {\bar z}_{31} {\bar z}_{32},
	\label{detMjm}
\end{equation}
to within a constant.  

Along the flow of one of the $I_r$'s we have ${\bar z}_{r\mu}(\psi_r)
= \exp(i\psi_r/2) {\bar z}_{r\mu}(0)$, or $\det M(\psi_r) =
\exp(i\psi_r) \det M(0)$.  Therefore when the given $\psi_r$ goes
from 0 to $4\pi$, the other $\psi_r$'s being held fixed (this is the
first basis contour), $\det M$ circles the origin twice and we have
$\mu=4$.  Along the flow of one of the $J_{rz}$'s, however, we have
${\bar z}_{r1}(\phi_r) = \exp(i\phi_r/2) {\bar z}_{r1}(0)$ and ${\bar
z}_{r2}(\phi_r) = \exp(-i\phi_r/2) {\bar z}_{r2}(0)$, or $\det
M(\phi_r) = \det M(0)$.  Therefore along the second basis contour the
$I_r$-flow takes $\det M$ once around the origin (elapsed parameter
$\psi_r=2\pi$), while the $J_{rz}$-flow does nothing.  Therefore the
Maslov index of the second basis contour is $\mu=2$.  There are easier
ways to find the Maslov indices of harmonic oscillators, but this
calculation is useful practice for the case of the Wigner manifold
that we take up momentarily.

Now we apply the quantization conditions.  For the first basis
contour the total phase is $4\pi j_r -4(\pi/2)$, which we set to
$2n_r\pi$ where $n_r$ is an integer.  Thus the quantized tori must
satisfy $j_r = (n_r+1)/2$.  The allowed values of $n_r$ are determined
by the fact $n_r < -1$ is impossible in view of the fact that $I_r$ is
nonnegative definite, and $n_r=-1$ corresponds to a torus of less than
full dimensionality (six), so the wave function (\ref{scpsi}) is not
meaningful.  Thus we must have $n_r=0,1,\ldots$.  The $j_r$ in these
formulas, and throughout all of the classical analysis from
Sec.~\ref{cmSchwinger} up to this point, has referred to a contour
value for the function $I_r$; the only difference now is that we are
restricting the value of $j_r$ in order that the torus be quantized.
This $j_r$, however, is not value of the principal symbol of the
operator ${\hat I}_r$ (see Eq.~(\ref{IJsymbols})), so the
Bohr-Sommerfeld or EBK quantization rule gives the semiclassical
eigenvalue of ${\hat I}_r$, call it $j_r^{\rm qu}$, as
\begin{equation}
	j_r^{\rm qu} = j_r-\frac{1}{2} = \frac{n_r}{2}.
	\label{jrquant}
\end{equation}
The semiclassical eigenvalues of ${\hat I}_r$ are nonnegative integers
or half-integers, the exact answer (not surprising in view of the fact
that semiclassical quantization of quadratic Hamiltonians is exact).
If we use the operator identity (\ref{JIidentity}) to find the
eigenvalues of operators ${\bf J}_r^2$, these are also exact.

Equation~(\ref{jrquant}) shows that the classical level set
corresponding to quantum number $j_r^{\rm qu}$ is $j_r = j_r^{\rm qu}
+ 1/2$.  The extra $1/2$ in this formula has caused some discussion in
the past and merits a little more now.  Ponzano and Regge (1968) used
intuition and numerical evidence to argue for the presence of the
$1/2$.  Miller, without knowing about Ponzano and Regge, also included
the $1/2$, referring to the ``usual'' semiclassical replacement for
angular momenta.  Presumably he was referring to the similar
replacement that occurs in the treatment of radial wave equations (the
Langer modification, see Berry and Mount (1972), Morehead (1995)).  It
is not obvious to us what the Langer modification has to do with the
$1/2$ that occurs in the present context, nor are we aware of any
general rules about when in the asymptotics of angular momentum theory
it is correct to replace a classical $j$ by $j+1/2$ (instead of
$[j(j+1)]^{1/2}$ or something else).  Schulten and Gordon (1975b) and
Reinsch and Morehead (1999) obtain the $1/2$ as a part of their proper
semiclassical analyses.  Biedenharn and Louck (1981b) also speculate
on the significance of the $1/2$.  Roberts' (1999) derivation of the
asymptotics of the $6j$-symbols does not produce the $1/2$.  He argues
that in the asymptotic limit there is no confusion about whether a
given point lies in the classically allowed or forbidden region,
whether the $1/2$ is included or not.  The omission of the $1/2$ does,
however, cause an error in the phase function that is of order unity,
so the oscillations are not even approximately represented (this point
was also made by Biedenharn and Louck, 1981b).  We suspect that a
modification of Roberts' method would produce the $1/2$'s.  According
to Girelli and Livine (2005), different choices for the semiclassical
replacement for the quantum number $j$ have been made by various
researchers in the field of quantum gravity.  Here we have shown that
the extra $1/2$ is a necessary consequence of standard semiclassical
theory.  We remark in addition that with the inclusion of the $1/2$,
the quantized spheres in angular momentum space are those with an area
of $(2j^{\rm qu}+1)2\pi$, that is, they contain a number of Planck
cells exactly equal to the dimension of the irrep, obviously a form of
geometric quantization.  In particular, the $s$-wave $j^{\rm qu}=0$ is
represented by a sphere of nonzero radius, a case for which the
replacement $j_r = j_r^{\rm qu} + 1/2$ is declared by Biedenharn and
Louck (1981b) to be ``clearly invalid.''

For the second basis contour on the $jm$-torus, the quantization
condition is $2\pi(j_r + m_r) - 2(\pi/2) = 2\pi n'_r$, where $n'_r$ is
an integer.  With (\ref{jrquant}), this implies $m_r = -j_r^{\rm
qu} + n'_r$.  Combined with the classical restriction
(\ref{jmranges}), this gives the usual range on magnetic quantum
numbers.  Again the semiclassical quantization is exact.  In the case
of $J_{rz}$, the eigenvalue of the operator is equal to the classical
contour value $m_r$ on the quantized torus (without any correction such as
we see in (\ref{jrquant})).

\subsection{Quantizing the Wigner manifold}

We begin the quantization of the Wigner manifold by guessing the basis
contours of the fundamental group by inspection of Fig.\ref{Wman}.
Taking the base (initial) point of the loops to be point $a$ in the
figure, we get three independent basis contours (call them $C_1$,
$C_2$, $C_3$) by going around the 3-torus $T_A$ in the three different
directions.  A fourth contour (call it $C_4$) is created by following
an $SU(2)$ rotation about some axis by angle $2\pi$, taking us along
the path $aba'$, which puts us half way around the torus $T_A$ from
the starting point, and then by applying half rotations along each of
the three directions on the torus, taking us down along $T_A$ in the
diagram back to the starting point $a$.  These four contours are not
independent, since
\begin{equation}
	2C_4 = C_1 + C_2 + C_3, 
	\label{homotopyrelation}
\end{equation}
where addition of contours means concatenation, but they are
convenient for studying the quanitzation conditions since a minimal
set of three contours (not $(C_1,C_2,C_3)$, but for example
$(C_1,C_2,C_4)$) is less symmetrical.  The fundamental group is
${\mathbb Z}^3$.

We may show the correctness of this guess by a topological argument.
First, the Wigner manifold is the orbit of a $SU(2) \times T^3$ group
action on the large phase space.  Let $(\psi_1,\psi_2,\psi_3)$ be
coordinates on $T^3$, where $0\le \psi_r \le 4\pi$.  The action of
$(u,(\psi_1, \psi_2, \psi_3)) \in SU(2) \times T^3$ on the large phase
space is to multiply each spinor by $u$ and then by
$\exp(-i\psi_r/2)$.  The isotropy subgroup of this action consists of the
identity $(1,(0,0,0))$ and the element $(-1,(2\pi,2\pi,2\pi))$, a
normal subgroup isomorphic to ${\mathbb Z}_2$.  Thus the Wigner
manifold is diffeomorphic to $(SU(2) \times T^3)/{\mathbb Z}_2$,
itself a group manifold, of which the original group $SU(2) \times
T^3$ is a double cover.  This cover is topologically simple since
$SU(2)$ is simply connected (we can go to the universal cover if we
wish by replacing $T^3$ by ${\mathbb R}^3$).  Therefore the homotopy
classes on the Wigner manifold are in one-to-one correspondence with
classes of topologically inequivalent curves that go from the identity
in $SU(2) \times T^3$ to one of the elements of the isotropy
subgroup.  Since $SU(2)$ is simply connected, such paths are
characterized by choice of the end point (the element of the isotropy
subgroup), and the winding numbers around the torus $T^3$.  They are
thus all ``linear combinations'' of the four contours $C_k, k=1,
\ldots, 4$ defined above.  Because of the relation
(\ref{homotopyrelation}), however, the fundamental group is not ${\mathbb
Z}^4$, but only ${\mathbb Z}^3$.

It is easy to compute the action along these contours.  Contours
$C_r$, $r=1,2,3$ follow the flows of the $I_r$'s, and the action is
the same as on the $jm$-torus, namely, $S_r = 4\pi j_r$.
Along contour $C_4$ the spinor rotation makes no contribution to the
action while the half rotation around the torus in all three angles
gives the action
\begin{equation}
	S_4 = 2\pi \sum_r j_r.
	\label{2Zaction}
\end{equation}

As for the Maslov indices, we compute the complex matrix of Poisson
brackets whose rows are indexed by the functions
$(I_1,I_2,I_3,J_x,J_y,J_z)$, and whose columns are indexed by $({\bar
z}_{11}, {\bar z}_{12}, {\bar z}_{21}, {\bar z}_{22}, {\bar z}_{31},
{\bar z}_{32})$.  To within a multiplicative constant that we drop, the
determinant is
\begin{eqnarray}
	&\left| \begin{array}{cccccc}
	{\bar z}_{11} & {\bar z}_{12} & 0 & 0 & 0 & 0 \\
	0 & 0 & {\bar z}_{21} & {\bar z}_{22} & 0 & 0 \\
	0 & 0 & 0 & 0 & {\bar z}_{31} & {\bar z}_{32} \\
	{\bar z}_{12} & {\bar z}_{11} & {\bar z}_{22}  &
	{\bar z}_{21} & {\bar z}_{32} & {\bar z}_{31} \\
	-{\bar z}_{12} & {\bar z}_{11} & -{\bar z}_{22} &
	{\bar z}_{21} & -{\bar z}_{32} & {\bar z}_{31} \\
	{\bar z}_{11} & -{\bar z}_{12} & {\bar z}_{21}  &
	-{\bar z}_{22} & {\bar z}_{31} & -{\bar z}_{32}
	\end{array}\right| \nonumber \\
	&={\rm const.} \times
	\left| \begin{array}{cc}
	{\bar z}_{11} & {\bar z}_{12} \\
	{\bar z}_{21} & {\bar z}_{22}
	\end{array}\right|
	\left| \begin{array}{cc}
	{\bar z}_{21} & {\bar z}_{22} \\
	{\bar z}_{31} & {\bar z}_{32}
	\end{array}\right|
	\left| \begin{array}{cc}
	{\bar z}_{31} & {\bar z}_{32} \\
	{\bar z}_{11} & {\bar z}_{12}
	\end{array}\right|.
	\label{2ZMi}
\end{eqnarray}
The final product of determinants is interesting, since these are the
$SU(2)$ invariants that Schwinger (Biedenharn and van Dam, 1965) used
to construct the rotationally invariant state $\vert j_1 j_2 j_3 {\bf
0} \rangle$ (with ${\bar z}_{r\mu}$ replaced by $a^\dagger_{r\mu}$).
Bargmann (1962) and Roberts (1999) make use of the same invariants.
For our purposes we need the winding number of the loop traced out in
the complex plane by the product of the three determinants as we
follow the four basis contours on the Wigner manifold.  

Proceeding as we did on the $jm$-torus, we find
that the Maslov indices along the contours $C_r, r=1,2,3$ are 4.
For example, along the $I_1$-flow ${\bar z}_{11}$ and ${\bar z}_{12}$
get multiplied by $\exp(i\psi_1/2)$, which causes two of the three
determinants to be multiplied by the same factor, so the product gets
multiplied by $\exp(i\psi_1)$, which has winding number 2 and hence
Maslov index 4 when $\psi_1$ goes from 0 to $4\pi$.  This is the same
answer we found along the $I_r$-flows on the $jm$-torus; this was not
exactly a foregone conclusion, even though the contours are the same,
because the tangent planes are different.  On the other hand, in both
cases we find the result (\ref{jrquant}) for the eigenvalues $j_r^{\rm
qu}$ of the operators ${\hat I}_r$, which of course must not depend on
how we compute them.  As for contour $C_4$ the first leg, a rotation
by angle $2\pi$ about some axis, leaves all three $2 \times 2$
determinants in (\ref{2ZMi}) invariant, so the big determinant in
the complex plane does not move.  As for the second leg, since each
$\psi_r$ only goes from 0 to $2\pi$ we get a winding number of 1 along
each $I_r$-flow, but since there are three of them the total winding
number is 3 and Maslov index is 6.

Combining this result with (\ref{2Zaction}), we obtain the
Bohr-Sommerfeld quantization condition for contour $C_4$
in the form
\begin{equation}
	2\pi \sum_r j_r - 6\frac{\pi}{2} = 2\pi \times {\rm integer},
	\label{2Zquant}
\end{equation}
or, with (\ref{jrquant}), 
\begin{equation}
	\sum_r j_r^{\rm qu} = {\rm integer}.
	\label{2zquant1}
\end{equation}
This is precisely the condition that the three quantum angular momenta
must satisfy, in addition to the triangle inequalities, that they may
add up to zero.  It emerges in a semiclassical analysis because the
Wigner manifold is not quantized otherwise.

In conclusion, the Bohr-Sommerfeld quantization conditions applied to
the $jm$-torus and the Wigner manifold give us a complete (and exact)
accounting of all the quantum numbers and the restrictions on them that
appear in the coupling of three angular momenta with a resultant of
zero.  It also allows us to identify the classical manifold (that is,
its contour values) with a given set of quantum numbers.  

\section{The amplitude determinant}
\label{ampdet}

The generic semiclassical eigenfunction of a complete set of commuting
observables is given by (\ref{scpsi}), with the amplitude
determinant expressed in terms of Poisson brackets by
(\ref{ampldet}).  These formulas apply in particular to the state
$\vert j_1 j_2 j_3 m_1 m_2 m_3\rangle$ on the left of the matrix
element (\ref{3jdef}), which is supported by the $jm$-torus in the
large phase space.  The state on the right, $\vert j_1 j_2 j_3 {\bf
0}\rangle$, however, which is supported by the Wigner manifold, is an
eigenfunction of observables that do not commute.  Therefore we must
rethink the derivation of Eqs.~(\ref{scpsi}) and (\ref{ampldet}) to
see what changes in this case.  In particular, we must see what
happens to the Poisson bracket expression for the amplitude
determinant, which is the solution of the simultaneous amplitude
transport equations for the collection of observables.  As it turns
out, nothing changes, the wave function is still given by
Eqs.~(\ref{scpsi}) and (\ref{ampldet}), with the (now noncommuting)
observables used in the amplitude determinant.  In addition, there is
a certain understanding about how the volume $V$ in (\ref{scpsi})
is computed, since the angles $\alpha$ conjugate to the $A$'s are no
longer meaningful.

Once this is done, we must evaluate the scalar product of the two wave
functions by stationary phase.  If both states were eigenstates of
complete sets of commuting observables, then the answer would be
(\ref{abmatrixelement1}) with amplitude determinant (\ref{PBdet}), but
again we must rethink the derivation of this result since the
observables for one of the wave functions do not commute.  Again, the
answer turns out to be given by formulas (\ref{abmatrixelement1}) and
(\ref{PBdet}) of Sec.~\ref{scwfis}, with a proper understanding of the
meanings of the volume factors.

Having established these facts, we can then proceed to the (easy)
calculation of the amplitude determinant for the $3j$-symbol in terms
of Poisson brackets, and finally put the remaining pieces together to
get the leading asymptotic form of the $3j$-symbols. 

\subsection{Amplitude determinant for noncommuting observables}

We begin showing that Eqs.~(\ref{scpsi}) and (\ref{ampldet}) are valid
for the state $\vert j_1 j_2 j_3 {\bf 0} \rangle$, with a proper
definition of the volume factors.  The classical functions defining
the Wigner manifold are $(I_1, I_2, I_3, J_x, J_y, J_z)$.  Let us
refer to these collectively as $A_k$, $k=1, \ldots, 6$, let us write
$x^i$, $i=1, \ldots, 6$ for the configuration space coordinates
instead of the notation $(x_{11}, \ldots, x_{32})$ used above, and let
us adopt the summation convention.  The functions $A_k$ form a Lie
algebra, that is, $\{A_k, A_l \} = c^m_{kl} \, A_m$, where $c^m_{kl}$
are the structure constants.  The Wigner manifold is a compact group
manifold with this Lie algebra, on which the Haar measure is both
left- and right-invariant.  This density is also invariant under the
flows generated by the right-invariant vector fields, which in our
case are the Hamiltonian flows of the functions $A_i$.  The projection
of this density onto configuration space is the density that provides
the solution of the simultaneous amplitude transport equations for the
functions $A_i$.  These are the basic geometrical facts, which we now
present more explicitly in coordinate language.

The amplitude transport equations for the functions $A_k$, $k=1,
\ldots, 6$ are
\begin{equation}
	\frac{\partial}{\partial x^i}
	\left[ \Omega(x) 
	\frac{\partial A_k}{\partial p_i}\right]=0,
	\label{ATeqn}
\end{equation}
where $p_i$ are the momenta conjugate to $x^i$.  These are six
simultaneous equations that must be solved for the density $\Omega(x)$
on configuration space.  Notice that $\partial A_k/\partial p_i =
\{x^i,A_k\} = {\dot x}^i_{(k)}$, the latter being notation we shall
use for the velocity in configuration space along the Hamiltonian flow
generated by $A_k$.  The amplitude transport equation is a continuity
equation, which is form-invariant under general coordinate
transformations.

Let us pick one of the branches of the inverse projection from
configuration space onto the Lagrangian (Wigner) manifold.  We shall
suppress the branch index in the following.  Let $u^i$, $i=1,\ldots,6$
be an arbitrary set of local coordinates on the Wigner manifold, which
we extend in a smooth but arbitrary manner into some small
neighborhood of the Wigner manifold, so that partial derivatives of
the $u^i$ with respect to all phase space coordinates are defined.
Assuming we are not at a caustic, the transformation from $x^i$ to
$u^i$ is locally one-to-one, and the Jacobian $\partial u^i/\partial
x^j$ is nonsingular.  Under the inverse projection or coordinate
transformation $x\to u$, the flow velocity transforms according to
\begin{equation}
	{\dot u}^i_{(k)} = 
	\frac{\partial u^i}{\partial x^j} \,
	{\dot x}^j_{(k)} = \{ u^i, A_k\} = X^i_{(k)},
	\label{velxfm}
\end{equation}
which defines the quantities $X^i_{(k)}$.  As a matrix, $X^i_{(k)}$ is
nonsingular because the flow vectors are linearly independent on the
Wigner manifold.  As for the density, it transforms according to
\begin{equation}
	\sigma(u) = \Omega(x) \left| \det
	\frac{\partial x^l}{\partial u^m}\right|,
	\label{densityxfm}
\end{equation}
so that the amplitude transport equations, lifted to the the Wigner
manifold, become
\begin{equation}
	\frac{\partial}{\partial u^i}
	[ \sigma(x) X^i_{(k)}]=0.
	\label{ATWm}
\end{equation}
Now define $\Lambda^{(k)}_j$ as the matrix inverse to $X^i_{(k)}$, 
\begin{equation}
	\Lambda^{(k)}_i X^i_{(l)} = \delta^k_l.
	\label{Lambdadef}
\end{equation}
As we will prove momentarily, the solution of Eqs.~(\ref{ATWm}) is 
\begin{equation}
	\sigma(u)= |\det \Lambda^{(k)}_j|,
	\label{sigmasoln}
\end{equation}
which, by (\ref{densityxfm}), gives us the solution of
(\ref{ATeqn}),
\begin{eqnarray}
	\Omega(x) &= \left| {\mathop{\rm det}\nolimits}_{kl} 
	\left(
	\Lambda^{(k)}_i 
	\frac{\partial u^i}{\partial x^l}\right)\right| =
	\left| {\mathop{\rm det}\nolimits}_{kl} 
	\left(
	X^i_{(k)} 
	\frac{\partial x^l}{\partial u^i}\right)\right|^{-1}
	\nonumber \\
	&= \left| {\mathop{\rm det}\nolimits}_{kl}\,
	\left({\dot u}^i_{(k)} 
	\frac{\partial x^l}{\partial u^i}\right)\right|^{-1}
	=\left| {\mathop{\rm det}\nolimits}_{kl}\,
	{\dot x}^l_{(k)} \right|=
	\left| {\mathop{\rm det}\nolimits}_{kl}\, 
	\{x^l,A_k \} \right|^{-1}.
	\label{Omegasoln}
\end{eqnarray}
In carrying out these manipulations it is important to note that
$\partial u^i/\partial x^j$ is taken at constant $A_k$, not $p_k$.
Thus the amplitude determinant for the wave function $\psi(x)$
associated with the Wigner manifold has the same Poisson bracket form
shown in (\ref{ampldet}), that is, in spite of the fact that the
$A_i$ do not commute.  

The essential differential geometry of these manipulations is that
$X_{(k)} = X^i_{(k)} \partial/\partial u^i$ are the Hamiltonian vector
fields on the Wigner manifold associated with functions $A_k$,
$\lambda^{(k)} = \Lambda^{(k)}_i \, du^i$ are the dual forms,
$\lambda^{(k)} X_{(l)} = \delta^k_l$, and $\sigma=\lambda^1 \wedge
\ldots \wedge \lambda^6 = \sigma(u) \,du^1 \wedge \ldots \wedge du^6$ 
is the Haar measure.  The condition (\ref{ATWm}) is equivalent to
$L_{X_{(k)}} \sigma = 0$.

To prove (\ref{sigmasoln}) in coordinates we substitute it into
(\ref{ATWm}) and expand out the derivative, obtaining an
expression proportional to
\begin{equation}
	X^i_{(k),i} - X^i_{(k)} X^l_{(m),i} \Lambda^{(m)}_l
	\label{intqty}
\end{equation}
using commas for derivatives.  Then we use the Lie bracket of the
vector fields $X_{(k)}$,
\begin{equation}
	[X_{(k)}, X_{(m)}]^l = X^i_{(k)} X^l_{(m),i} -
	X^i_{(m)} X^l_{(k),i} = -c^n_{km} X^l_{(n)},
	\label{Liebracket}
\end{equation}
where $c^n_{km}$ are the structure constants.  Here we use the
identity expressing the Lie bracket of Hamiltonian vector fields for
two functions in terms of the Hamiltonian vector field of their
Poisson bracket, $[X_H, X_K] = -X_{\{H,K\}}$ (Arnold, 1989).  Thus
(\ref{intqty}) becomes simply $c^m_{km}$, which vanishes since for the
group in question the structure constants are completely
antisymmetric.

Finally to normalize the semiclassical eigenfunction supported by the
Wigner manifold we use the stationary phase approximation to compute
the integral
\begin{equation}
	\int dx \, \left| \sum_{\rm br} \Omega(x)^{1/2}
	\exp[iS(x)-i\mu\pi/2]\right|^2,
	\label{Wignorm}
\end{equation}
where the sum is over branches and the branch index is suppressed.
Cross terms do not contribute, and when the integral is lifted to the
Wigner manifold it just gives the volume of that manifold with respect
to the Haar measure,
\begin{equation}	
	\sum_{\rm br} \int \frac{dx}
	{|{\mathop{\rm det}\nolimits} \{x^i, A_j \}|}
	= \int du\,\sigma(u) = V_W,
	\label{Wignorm1}
\end{equation}
where $V_W$ is given by (\ref{Wignervolume}).

\subsection{Matrix elements for noncommuting observables}

Now we write the $3j$ matrix element (\ref{3jdef}) as $\langle b \vert
a \rangle$, where $A=(I_1, I_2, I_3, J_x, J_y, J_z)$ and $B=(I_1, I_2, I_3,
J_{1z}, J_{2z}, J_{3z})$.  Actually this is not the most convenient
form, since $J_z$ in the $A$-list is a function of the $J_{rz}$ in the
$B$-list.  We fix this by performing a canonical transformation
$(\phi_1, \phi_2, \phi_3, J_{1z}, J_{2z}, J_{3z}) \to ({\tilde
\phi}_1, {\tilde \phi}_2, {\tilde \phi}_3, {\tilde J}_{1z}, {\tilde
J}_{2z}, J_z)$ on the functions in the $B$-list, generated by
\begin{equation}
	\fl F_2(\phi_1,\phi_2,\phi_3, {\tilde J}_{1z}, {\tilde J}_{2z},
	J_z) = \phi_1 {\tilde J}_{1z} + \phi_2 {\tilde J}_{2z}
	+ \phi_3(J_z - {\tilde J}_{1z} - {\tilde J}_{2z}).
	\label{JzCT} 
\end{equation}
This gives ${\tilde J}_{1z} = J_{1z}$, ${\tilde J}_{2z} = J_{2z}$,
$J_z = J_{1z} + J_{2z} + J_{3z}$, and ${\tilde \phi}_1 =
\phi_1 - \phi_3$, ${\tilde \phi}_2 = \phi_2-\phi_3$, ${\tilde \phi}_3
= \phi_3$.  The linear transformation in the angles has unit
determinant, so the volume of the $jm$-torus is still given by
(\ref{Vjm}).  Dropping the tildes, the $B$-list is now
$(I_1,I_2,I_3, J_{1z}, J_{2z}, J_z)$, which has four functions in
common with the $A$-list.

Now the integral we must evaluate is
\begin{eqnarray}
	\langle b \vert a \rangle &=
	\frac{1}{\sqrt{V_A V_B}} \int dx
	\frac{1}{|{\mathop {\rm det}} \{ x^i, A_j \}|^{1/2}}
	\frac{1}{|{\mathop {\rm det}} \{ x^i, B_j \}|^{1/2}}
	\nonumber \\
	&\qquad \times
	\sum_{\rm br}
	\exp\{i[S_A(x)-S_B(x) - \mu\pi/2]\},
	\label{abintegral}
\end{eqnarray}
where the sum is over all branches of the projections of the two
manifolds, and where $\mu$ just stands for whatever Maslov index
appears in a given term (different $\mu$'s are not necessarily
equal).  An integral like this was evaluated by Littlejohn (1990),
using the angles conjugate to the $A$'s and $B$'s, but those do not
all exist in the present circumstances and we must evaluate the
integral in a different way.  

Let us write $A=(C,D)$ and $B=(C,E)$, where $C=(I_1, I_2, I_3, J_z)$
are the four observables in common in the $A$- and $B$-lists, and
where $D=(J_x, J_y)$ and $E=(J_{1z}, J_{2z})$ are the two pairs of
observables that are distinct.  The stationary phase set of the
integral (\ref{abintegral}) consists of points $x$ where
$\partial(S_A-S_B)/\partial x^i=0$, that is, it is the projection onto
configuration space of the intersection of the $A$-manifold and the
$B$-manifold.  That intersection, which we denote by $I$, was
studied in Sec.~\ref{intersections} (it is a 4-torus).  It is the
simultaneous level set of all of the $A$'s and $B$'s, and at the same
time the orbit of the commuting Hamiltonian flows generated by the
$C$'s.  Its projection onto configuration space is a 4-dimensional
region.

We introduce a local coordinate transformation in configuration space
$x \to (y,z)$ where the four $y$'s are coordinates along the stationary
phase set and the two $z$'s are transverse to it.  We let the stationary
phase set itself be specified by $z=0$.  We let $(u,v)$ be the momenta
conjugate to $(y,z)$. Then the two amplitude determinants in
(\ref{abintegral}) may be combined with the Jacobian of
the coordinate transformation to result in the square root of the
product of two determinants, one of which is
\begin{equation}
	\det \frac{\partial(y,z)}{\partial x}
	\det \{ x, A \} =
	\det \left(\begin{array}{cc}
	\{ y, C \} & \{ y, D \} \\
	\{ z, C \} & \{ z, D \}
	\end{array}
	\right),
	\label{yzCDPB}
\end{equation}
and the other of which is the same but with the substitutions $A\to
B$, $D \to E$.  But since the $C$'s generate flows along $I$, we have
$\{z^i,C_j\}=0$, and the lower left block of the two matrices
vanishes.  Thus, the product of the two determinants becomes
\begin{equation}
	[\det \{ y, C \}]^2 \det \{ z, D\} \det \{ z, E\},
	\label{detprod}
\end{equation}
the square root of which appears in the denominator of the integrand.
Evaluating the final two Poisson brackets in the $(y,z;u,v)$ canonical
coordinates, we have
\begin{equation}
	\{z^i, D_j \} = \frac{\partial D_j}{\partial v_i},
	\qquad
	\{z^i, E_j \} = \frac{\partial E_j}{\partial v_i}.
	\label{zDzEPBs}
\end{equation}

We perform the $z$-integration by stationary phase, expanding $S_A$
and $S_B$, regarded as functions of $(y,z)$, to second order in $z$
for a fixed value of $y$, and simply evaluating the amplitude at $z=0$
(that is, on $I$).  To within a phase, the $z$-integration gives
\begin{equation}
	2\pi
	\left| \det \left(
	\frac{\partial^2 S_A}{\partial z\partial z}
	-\frac{\partial^2 S_B}{\partial z\partial z}\right) 
	\right|^{-1/2}.
	\label{zintegral}
\end{equation}
The determinant in this result must be multiplied by the determinants
of the matrices (\ref{zDzEPBs}) to get the overall determinant in the
denominator after the $z$-integration.  The product of these three
determinants is the determinant of the matrix
\begin{equation}
	\frac{\partial D}{\partial v}\left[
	\left(\frac{\partial v}{\partial z}\right)_{yCD} -
	\left(\frac{\partial v}{\partial z}\right)_{yCE}\right]
	\left(\frac{\partial E}{\partial v}\right)^T,
	\label{3matrixprod}
\end{equation}
where it is understood that a partial derivative stands for a matrix
whose row index is given by the numerator and column index by the
denominator, unless the matrix transpose or inverse is indicated, in
which case the rule is reversed.  Also, if a partial derivative is
shown without subscripts, then it is assumed that it is computed in
the canonical coordinates $(y,z;u,v)$, and otherwise the variables to
be held fixed are explicitly indicated.  In the two middle matrices in
(\ref{3matrixprod}), the variables held fixed amount to
differentiating $v$ with respect to $z$ along the $A$- and
$B$-manifolds, respectively, since $\partial S_A/\partial z = v(x,A)$
and $\partial S_B/\partial z=v(x,B)$.   Notice that these two matrices
are symmetric.
	
Now we express the two matrices in the middle of
(\ref{3matrixprod}) purely in terms of partial derivatives
computed in the canonical $(y,z;u,v)$ coordinates.  We do this by
writing out the Jacobian matrix $\partial(y,z;C,D)/ \partial(y,z;u,v)$
and the inverse Jacobian $\partial(y,z;u,v)/\partial(y,z;C,D)$,
multiplying the two together to obtain a series of identities
connecting the forward and inverse Jacobian blocks, and then solving
for the inverse Jacobian blocks in terms of the forward ones.  We note
that the block $\partial C/\partial v$ of the forward Jacobian
vanishes, since it is $\{z,C\}$.  Thus we find
\begin{equation}
	\fl\eqalign{
	\left(\frac{\partial v}{\partial z}\right)_{yCD} &=
	\left(\frac{\partial D}{\partial v}\right)^{-1} \left[
	\frac{\partial D}{\partial u}
	\left(\frac{\partial C}{\partial u}\right)^{-1}
	\frac{\partial C}{\partial z}-
	\frac{\partial D}{\partial z}\right], \cr
	\left(\frac{\partial v}{\partial z}\right)_{yCE} &=
	\left[\left(\frac{\partial C}{\partial z}\right)^T
	\left(\frac{\partial C}{\partial u}\right)^{-1T}
	\left(\frac{\partial E}{\partial u}\right)^T -
	\left(\frac{\partial E}{\partial z}\right)^T\right]
	\left(\frac{\partial E}{\partial v}\right)^{-1T}. \cr}
	\label{vzidents}
\end{equation}
Upon substituting these into (\ref{3matrixprod}), that matrix
becomes
\begin{eqnarray}
	&\frac{\partial D}{\partial v}
	\left(\frac{\partial E}{\partial z}\right)^T -
	\frac{\partial D}{\partial z}
	\left(\frac{\partial E}{\partial v}\right)^T 
	+\frac{\partial D}{\partial u}
	\left(\frac{\partial C}{\partial u}\right)^{-1}
	\frac{\partial C}{\partial z}
	\left(\frac{\partial E}{\partial v}\right)^{T}
	\nonumber \\
	&-\frac{\partial D}{\partial v}
	\left(\frac{\partial C}{\partial z}\right)^T
	\left(\frac{\partial C}{\partial u}\right)^{-1T}
	\left(\frac{\partial E}{\partial u}\right)^T,
	\label{firsthalfDEPB}
\end{eqnarray}
where the first two terms are the beginning of the Poisson bracket
$\{E,D\}$.  As for the last two terms, we write out the vanishing
Poisson brackets $\{C,E\}$ and $\{C,D\}$ in the $(y,z;u,v)$
coordinates, making use of $\partial C/\partial v=0$, to obtain
\begin{equation}
	\eqalign{
	\frac{\partial C}{\partial z}
	\left(\frac{\partial E}{\partial v}\right)^T
	&= \frac{\partial C}{\partial u}
	\left(\frac{\partial E}{\partial y}\right)^T -
	\frac{\partial C}{\partial y}
	\left(\frac{\partial E}{\partial u}\right)^T, \cr
	\frac{\partial D}{\partial v}
	\left(\frac{\partial C}{\partial z}\right)^T
	&= \frac{\partial D}{\partial y}
	\left(\frac{\partial C}{\partial u}\right)^T -
	\frac{\partial D}{\partial u}
	\left(\frac{\partial C}{\partial y}\right)^T. \cr}
	\label{CEPBident}
\end{equation}
Actually the matrix of Poisson brackets $\{C,D\}$ does not vanish
everywhere in phase space, just on the $A$- (or Wigner) manifold, and
in particular on the intersection $I$ which is where we are evaluating
them.  Now substituting Eqs.~(\ref{CEPBident}) into the last two terms
of (\ref{firsthalfDEPB}), those terms become
\begin{eqnarray}
	&\frac{\partial D}{\partial u}
	\left(\frac{\partial E}{\partial y}\right)^T -
	\frac{\partial D}{\partial y}
	\left(\frac{\partial E}{\partial u}\right)^T \\
	&+\frac{\partial D}{\partial u} \left[
	\left(\frac{\partial C}{\partial y}\right)^T
	\left(\frac{\partial C}{\partial u}\right)^{-1T} -
	\left(\frac{\partial C}{\partial u}\right)^{-1}
	\frac{\partial C}{\partial y}\right]
	\left(\frac{\partial E}{\partial u}\right)^T,
	\label{secondhalfDEPB}
\end{eqnarray}
in which the first two terms give us the remainder of the Poisson
bracket $\{E,D\}$.  As for the last major term, the factor in the
square brackets vanishes, as we see by writing out the vanishing
Poisson bracket $\{C,C\}$ in coordinates $(y,z;u,v)$ and using
$\partial C/\partial v=0$.  

As a result the integral (\ref{abintegral}) becomes
\begin{equation}
	\langle b \vert a \rangle =
	\frac{2\pi}{\sqrt{V_A V_B}}
	\sum_{\rm br}
	\int \frac{dy}{|\det\{y,C\}|}
	\frac{e^{i(S_I - \mu\pi/2)}}
	{|\det\{E,D\}|^{1/2}},
	\label{abyintegral}
\end{equation}
where the branch sum runs over all branches of the projection of $I$
onto configuration space as well as the two disconnected components of $I$
(the two 4-tori discussed in Sec.~\ref{intersections}), and where
$S_I$ is the phase on a given connected component of $I$ (this is the
phase $\pm S_{jm}$ computed in Sec.~\ref{actionintegrals}).  We have also
dropped an overall phase, and we are not attempting to compute the
Maslov indices in detail.  The amplitude determinant has been reduced
to a $2\times 2$ matrix of Poisson brackets of the observables in the
$A$- and $B$-lists that differ, exactly as in (\ref{PBdet}).
Calculating this matrix explicitly, we find 
\begin{eqnarray}
	|\det\{E,D\}| &= \left|\begin{array}{cc}
	\{ J_{1z}, J_x \} & \{ J_{1z}, J_y \} \\
	\{ J_{2z}, J_x \} & \{ J_{2z}, J_y \}
	\end{array}\right| =
	\left|\begin{array}{cc}
	J_{y1} & -J_{x1} \\
	J_{y2} & -J_{x2} 
	\end{array}\right|
	\nonumber \\
	[3pt]
	&= |J_{x1} J_{y2} - J_{x2} J_{y1}|
	= | {\bf z} \cdot ({\bf J}_1 \times {\bf J}_2)|
	= 2 \Delta_z,
	\label{3jampldet}
\end{eqnarray}
where $\Delta_z$ is the projection of the area of the triangle
$\Delta$ onto the $x$-$y$ plane (see Eq.~(\ref{Deltadef}) and Fig.~2
of Ponzano and Regge (1968)).  This quantity is invariant under
rotations about the $z$-axis, that is, it Poisson commutes with $J_z$.
It also Poisson commutes with the other three variables in the
$C$-list, $(I_1,I_2,I_3)$, and so is constant on the intersection
$I$ and can be taken out of the $y$-integral.  The same applies to the
phase factor, since $S_I$ is also constant on the $I$-manifold.  Then
the $y$-integral can be done, since $|\det\{y,C\}|$ is just the
Jacobian connecting $y$ with the angle variables conjugate to
$C=(I_1,I_2,I_3,J_z)$, denoted above by $(\psi_1,\psi_2,\psi_3,\phi)$.
Thus the $y$-integral just gives the volume $V_I$ of the intersection $I$
with respect to these angles, see (\ref{VIdef}).  In fact, had the
variables $C$ not been commuting, but if they had formed a Lie
algebra, then $V_I$ would be the volume of $I$ with respect to the
Haar measure of the corresponding group.  This circumstance arises,
for example, in a similar treatment of the $6j$-symbol.

As a result of these rather lengthy manipulations of amplitude
determinants, we obtain the final, simple result,
\begin{equation}
	\langle b \vert a \rangle =
	\frac{2\pi}{\sqrt{V_A V_B}} \sum_{\rm br} V_I
	\frac{e^{i(S_I-\mu\pi/2)}}{|\det\{E,D\}|^{1/2}},
	\label{integraldone}
\end{equation} 
where the branches now run over just the two disconnected pieces of
the intersection $I$.  This is a version of
(\ref{abmatrixelement1}), with the right understanding of the
volume measures, generalized to the case at hand in which the
observables do not commute.  The actual calculation of the final amplitude
determinant takes just one line, Eq.~(\ref{3jampldet}).  

In fact, for our application the volume $V_I$ and the remaining
amplitude determinant are the same for both branches and can be taken
out of the sum.  The relative Maslov index between the two branches is
1; we will not belabor this point since the answer is already known.
We simply note that by splitting the Maslov phase $i\pi/2$ between
the two branches and subsitituting $V_A=V_W$, $V_B=V_{jm}$, we obtain
to within an overall phase the result of Ponzano and Regge,
\begin{equation}
	\left(
  \begin{array}{ccc}
    j_1 & j_2 & j_3 \\
    m_1 & m_2 & m_3 
  \end{array}
  \right) = ({\rm phase}) \times 
	\frac{\cos(S_{jm}+\pi/4)}{\sqrt{2\pi\Delta_z}}.
	\label{PRresult}
\end{equation}

\section{Conclusions}
\label{conclusions}

In many ways the $3j$-symbol is not as interesting as the $6j$-symbol,
of which it is a limiting case.  We intended our work on the
$3j$-symbol as a warm-up exercise, expecting a routine application of
semiclassical methods for integrable systems.  The nongeneric
Lagrangian (Wigner) manifold was a surprise.  Similar nongeneric
Lagrangian manifolds occur also in the semiclassical analysis of the
$6j$- and $9j$-symbols.

If all one wants is a derivation of an asymptotic formula, then there
are many ways to proceed.  For example, one can simply take the
expression for the symbol due to Wigner ($3j$) or Racah ($6j$) as a
sum over a single index, and apply standard asymptotic methods
(Stirling's approximation, Poisson sum rule, etc).  But if one wants a
derivation that reveals the geometrical meaning of the classical
objects that emerge (the triangle, the tetrahedron, etc), then an
approach such as ours may be preferable.

Our approach is more geometrical than earlier ones, and in that
respect is closer in spirit to the work of Roberts (1999), Freidel and
Louapre (2003) and later authors.  It is likely that at some deeper
level all these methods are the same, although superficially we see
only a little similarity between our work and these others.  

One may also desire a method that makes the symmetries of the symbol
manifest.  Our analysis does not do this for the $3j$-symbol, but
those symmetries are not manifest in Wigner's definition of the
$3j$-symbol that we employ as our starting point, either.  To bring
the symmetries out it seems necessary to employ some construction
related to Schwinger's generating functions, which involve lifting the
definitions into higher dimensional spaces.

Our method of calculating amplitude determinants in terms of Poisson
brackets may have computational advantages in other applications, as
well.  The method can be remarkably easy to use.  For example, the
$6j$-symbol can be defined as a matrix element,
\begin{equation}
	\left\{ \begin{array}{ccc}
	j_1 & j_2 & j_{12} \cr
	j_4 & j_3 & j_{23} 
	\end{array}\right\}
	= {\rm const.} \times \langle j_1j_2j_3j_4 j_{23} {\bf 0} \vert
          j_1 j_2 j_3 j_4 j_{12} {\bf 0} \rangle,
	\label{6jdef}
\end{equation}
which is the unitary matrix in $(j_{12},j_{23})$ defining a change of
basis in the subspace in which four angular momenta of given lengths
add up to zero (${\bf 0}$ means ${\bf J}={\bf 0}$).  In this case
there are eight observables on each side of the matrix element, of
which seven are common and one is different.  Thus the amplitude of
the $6j$-symbol is the inverse square root of the single Poisson
bracket,
\begin{equation}
	\{{\bf J}_{23}^2, {\bf J}_{12}^2\} = 4
	{\bf J}_1 \cdot ({\bf J}_2 \times {\bf J}_3),
	\label{6jPB}
\end{equation}
as follows immediately from (\ref{NjLiePoisson}).  One sees
immediately that it is proportional to the volume of the tetrahedron.
A similarly easy calculation is possible for the $9j$-symbol.  It is
harder, however, to express these amplitudes in terms of the quantum
numbers (the magnitudes $j_r$), that is, to translate these magnitudes
into vectors ${\bf J}_r$ that lie on the stationary phase set.  We
shall report on these and other extensions of our work in future
publications.

\section*{References}
\begin{harvard}

\item[] Anderson R W and Aquilanti V 2006 {\it J. Chem. Phys.} {\bf
124} 214104

\item[] Aquilanti V, Cavalli S and De~Fazio D 1995 {\it J. Phys. Chem.}
{\bf 99} 15694

\item[] Aquilanti V, Cavalli S and Coletti C 2001 {\it Chem. Phys. Lett.}
{\bf 344} 587

\item[] Aquilanti V and Coletti C 2001 {\it Chem. Phys. Lett.} {\bf
344} 601 

\item[] Arnold V I 1967 {\it Functional Anal.\ Appl.} {\bf 1} 1

\item[] \dash 1989 {\it Mathematical Methods of Classical
Mechanics} (New York: Springer-Verlag)

\item[] Barrett J W and Steele C M 2002 {\it Class. Quant. Grav.} {\bf
20} 1341

\item[] Baez J C, Christensen J D and Egan G 2002 {\it
Class. Quant. Grav.} {\bf 19} 6489

\item[] Balazs N L and Jennings B K 1984 {\it Phys. Reports} {\bf 104}
347

\item[] Bargmann V 1962 {\it Rev. Mod. Phys.} {\bf 34} 829

\item[] Berry M V 1977 {\it Phil. Trans. Roy. Soc.} {\bf 287} 237

\item[] Berry M V and Mount K E 1972 {\it Rep.\ Prog.\ Phys.} {\bf 35}
315

\item[] Berry M V and Tabor M 1976 {\it Proc.\ Roy.\ Soc.\ Lond.\ A}
{\bf 349} 101

\item[] Biedenharn L C and Louck J D 1981a {\it Angular Momentum in 
Quantum Physics} (Reading, Massachusetts: Addison-Wesley)

\item[] \dash 1981b {\it The Racah-Wigner Algebra in Quantum Theory}
(Reading, Massachusetts: Addison-Wesley)

\item[] Biedenharn L C and van Dam H 1965 {\it Quantum Theory of
  Angular Momentum} (New York: Academic Press)

\item[] Brack Matthias and Bhaduri Rajat K 1997 {\it Semiclassical
Physics} (Reading, Massachusetts: Addison-Wesley)

\item[] Brillouin M L 1926 {\it J.\ Phys.} {\bf 7} 353

\item[] Cargo Matthew, Gracia-Saz Alfonso, Littlejohn R  G , Reinsch
  M W and de M. Rios P 2005,  {\it J. Phys.\ A} {\bf 38} 1977

\item[] Cargo Matthew, Gracia-Saz Alfonso and Littlejohn R G 2005,
  preprint math-ph/0507032.

\item[] Cushman R H and Bates L 1997 {\it Global Aspects of Classical 
Integrable Systems} (Basel: Birkh\"auser Verlag)

\item[] De~Fazio D, Cavalli S and Aquilanti V 2003 {\it
Int. J. Quant. Chem.} {\bf 93} 91

\item[] de Gosson, M 1997 {\it Maslov Classes, Metaplectic
Representation and Lagrangian Quantization (Mathematical Research
Vol. 5)} (Berlin: Akademischer Verlag)

\item[] Einstein A 1917 {\it Verh.\ dt.\ Phys.\ Ges.} {\bf 19} 82

\item[] Estrada Ricardo, Gracia-Bond\'\i a J M and V\'arilly J C 1989
{\it J. Math. Phys.} {\bf 30} 2789 

\item[] Frankel Theodore 1997 {\it The Geometry of Physics} (Cambridge)

\item[] Freidel Laurent and Louapre David 2003 {\it Class. Quantum Grav.}
{\bf 20} 1267

\item[] Geronimo J S, Bruno O and Van Assche W 2004 {\it Oper. Theory
Adv. Appl.} {\bf 154} 101

\item[] Girelli F and Livine E R 2005 {\it Class. Quantum Grav.} 
{\bf 22} 3295

\item[] Gracia-Bond\'\i a J M and V\'arilly J C 1995 {\it
J. Math. Phys.} {\bf 36} 2691

\item[] Groenewold H J 1946 {\it Physica} {\bf 12} 405

\item[] Gutzwiller Martin C 1990 {\it Chaos in Classical and Quantum
Mechanics} (New York: Springer-Verlag)

\item[] Hillery M, O'Connell R F, Scully M O and Wigner E P 1984 {\it
Phys. Reports} {\bf 106} 123

\item[] Keller J B 1958 {\it Ann.\ Phys.} {\bf 4} 180

\item[] Littlejohn R G 1986 {\it Phys. Reports} {\bf 138} 193

\item[] \dash 1990 {\it  J. Math.\ Phys.} {\bf 31} 2952

\item[] Littlejohn R G and Reinsch M 1995 {\it Phys.\ Rev.\ A} {\bf52}
2035

\item[] Littlejohn R G and Robbins J M 1987 {\it Phys.\ Rev.\ A}
{\bf 36} 2953
 
\item[] Marsden J E and Ratiu T 1999 {\it Introduction to Mechanics and
  Symmetry} (New York: Springer-Verlag)

\item[] Marzuoli A and Rasetti M 2005 {\it Ann. Phys.} {\bf 318} 345

\item[] Maslov V P and Fedoriuk M V 1981 {\it Semi-Classical
Approximations in Quantum Mechanics} (Dordrecht: D. Reidel)

\item[] McDonald S 1988 {\it Phys. Reports} {\bf 158} 377

\item[] Miller W H 1974 {\it Adv. Chem. Phys.} {\bf 25} 69

\item[] Mishchenko A S, Shatalov V E and Sternin B Yu 1990 {\it Lagrangian
manifolds and the Maslov operator} (Berlin: Springer Verlag)

\item[] Morehead J J 1995 {\it J.\ Math.\ Phys.} {\bf 36} 5431

\item[] Moyal J E 1949 {\it Proc. Camb. Phil. Soc.} {\bf 45} 99

\item[] Nakahara M 2003 {\it Geometry, Topology and Physics} (Bristol:
  IOP Publishing)

\item[] Neville Donald 1971 {\it J. Math. Phys.} {\bf 12} 2438

\item[] Ozorio de Almeida Alfredo M 1998 {\it Phys.\ Reports} {\bf
295} 265

\item[] Percival I C 1973 {\it J.\ Phys.\ B} {\bf 6} L229

\item[] Ponzano G and Regge T 1968 in {\it Spectroscopy and Group
Theoretical Methods in Physics} ed F Bloch \etal\ (Amsterdam:
North-Holland) p~1

\item[] Reinsch M W and Morehead J J 1999 {\it J.\ Math.\ Phys.} {\bf
40} 4782

\item[] Roberts J 1999 {\it Geometry and Topology} {\bf 3} 21

\item[] Sakurai J J 1994 {\it Modern Quantum Mechanics} (New York:
  Addison-Wesley)

\item[] Schulman L S 1981 {\it Techniques and Applications of Path
  Integration} (New York:  John Wiley \& Sons)

\item[] Schulten K and Gordon R G 1975a {\it J. Math. Phys.} {\bf 16} 1961

\item[] \dash 1975b {\it J. Math. Phys.} {\bf 16} 1971

\item[] Smorodinskii Ya A and Shelepin L A 1972 {\it Sov. Phys. Usp.}
  {\bf 15} 1

\item[] Taylor Y U and Woodward C T 2005 {\it Selecta Math. (N.S.)}
{\bf 11} 539

\item[] Voros A 1977 {\it Ann. Inst. Henri Poincar\'e} {\bf 4} 343

\item[] Weyl H 1927 {\it Z. Phys.} {\bf 46} 1

\item[] Wigner E P 1932 {\it Phys. Rev.} {\bf 40} 749

\item[] \dash 1959 {\it Group Theory} (Academic Press, New York)

\end{harvard}

\Figures

\begin{figure}
\caption{\label{SU2action} The action of an $SU(2)$ rotation about a
  fixed axis on a point of the large phase space, and its projection
  onto angular momentum space.}
\end{figure} 

\begin{figure}
\caption{\label{xpplane} Harmonic oscillator motion in the
  $x_{r\mu}$-$p_{r\mu}$ plane, generated by $I_{r\mu}$.  Definition of
  angle $\theta_{r\mu}$ is shown.}
\end{figure}

\begin{figure}
\caption{\label{3Jcones} The quantum state $\vert j_1j_2j_3 m_1m_2m_3
  \rangle$ corresponds classically to a 3-torus in the small phase
  space $S^2 \times S^2 \times S^2$, which can be visualized as three
  ${\bf J}$ vectors lying on three cones with fixed values of $J_z$.  The
  three azimuthal angles are independent.}
\end{figure}

\begin{figure}
\caption{\label{Jtriangle} If $j_1$, $j_2$ and $j_3$ satisfy the
  triangle inequalities, then they define a triangle that is unique
  apart from its orientation.  A standard orientation places the
  triangle in the $x$-$z$ plane with sides ${\bf J}_1$, ${\bf J}_2$,
  ${\bf J}_3$ oriented as shown.  The angle opposite ${\bf J}_r$ is
  $\eta_r$.}
\end{figure}

\begin{figure}
\caption{\label{Wman} An schematic illustration showing how the Wigner
  manifold in the large phase space is the inverse image under $\pi$
  of a set of triangles formed from three angular momentum vectors
  with vanishing sum, all related by rigid rotations.}
\end{figure}

\begin{figure}
\caption{\label{beta} By rotating the reference orientation of the
  triangle about the $y$-axis, we can give ${\bf J}_3$ the desired
  projection $m_3$ onto the $z$-axis.}
\end{figure}

\begin{figure}
\caption{\label{gamma} Once vector ${\bf J_3}$ has the desired
  projection $m_3$, we rotate the triangle by angle $\gamma$ about the
  axis ${\bf J}_3$ to make ${\bf J}_2$ have its desired projection
  $m_2$.  This cannot always be done for real angles $\gamma$, but
  when it can be done there are generically two angles that work,
  illustated by points $Q$ and $Q'$ in the figure.}
\end{figure}

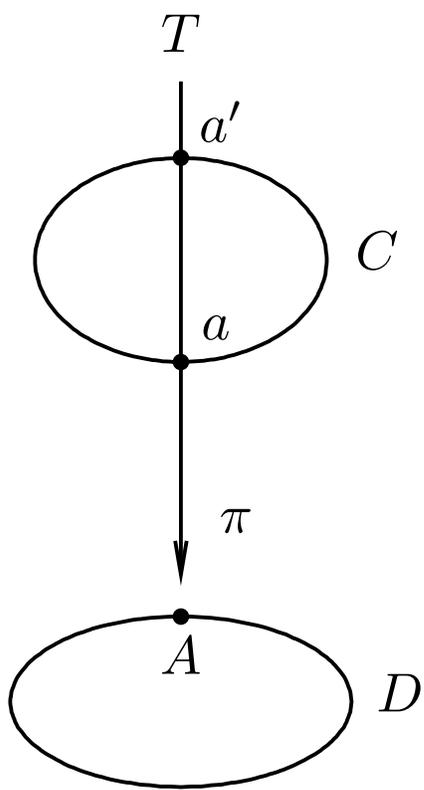
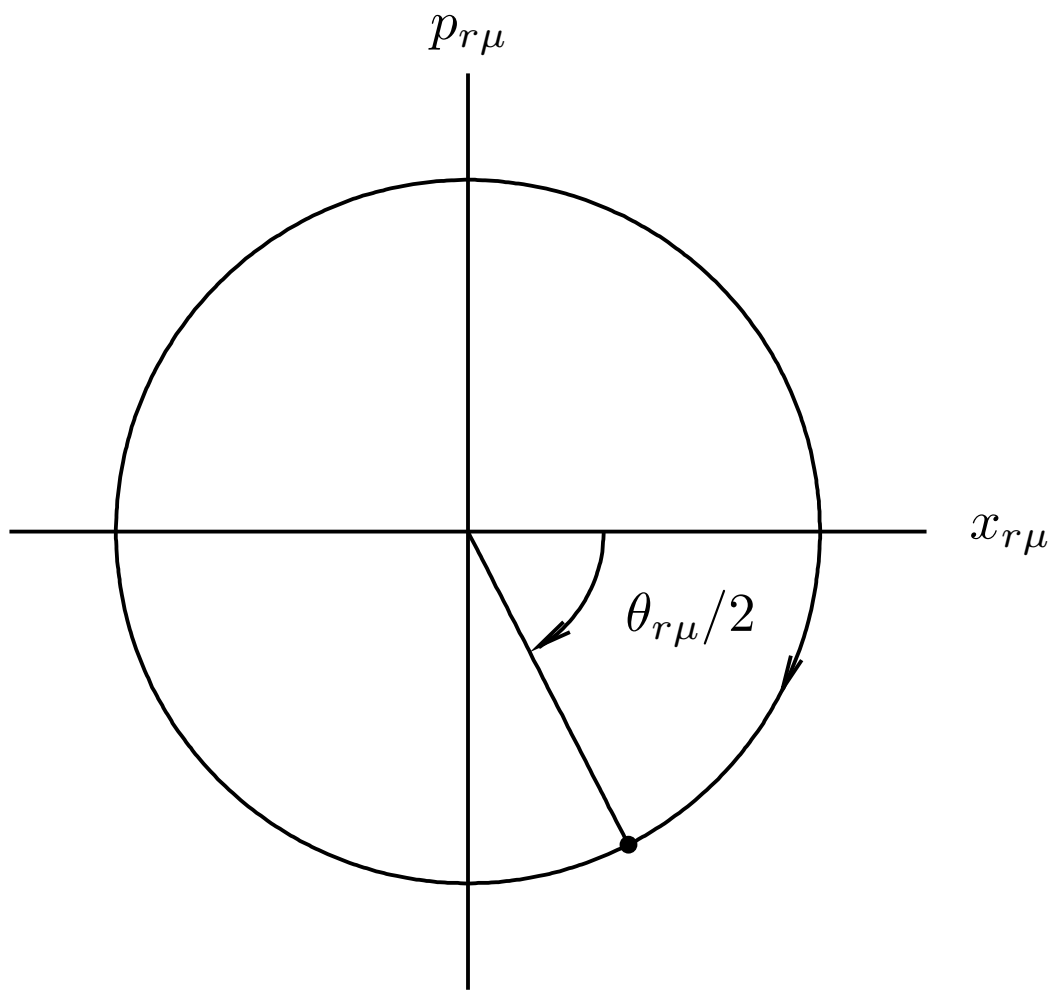
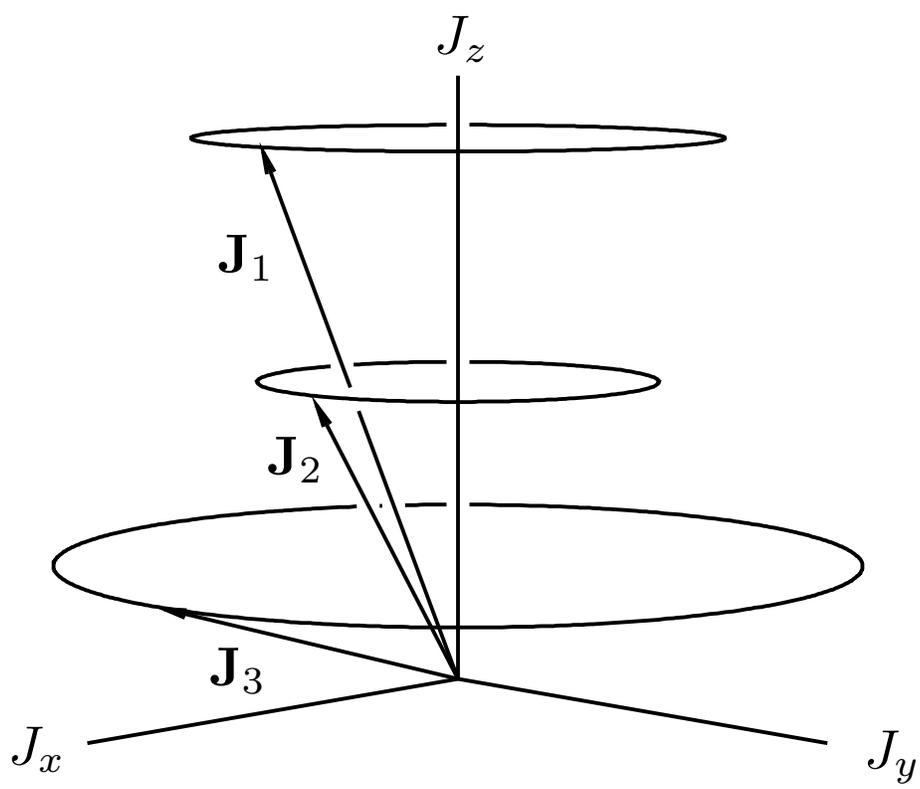
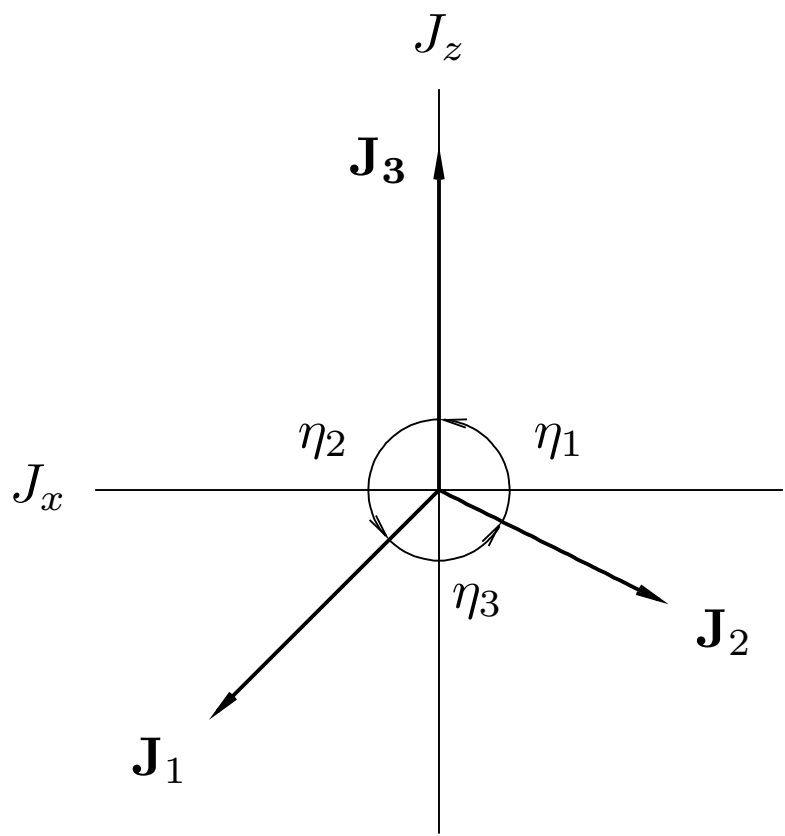
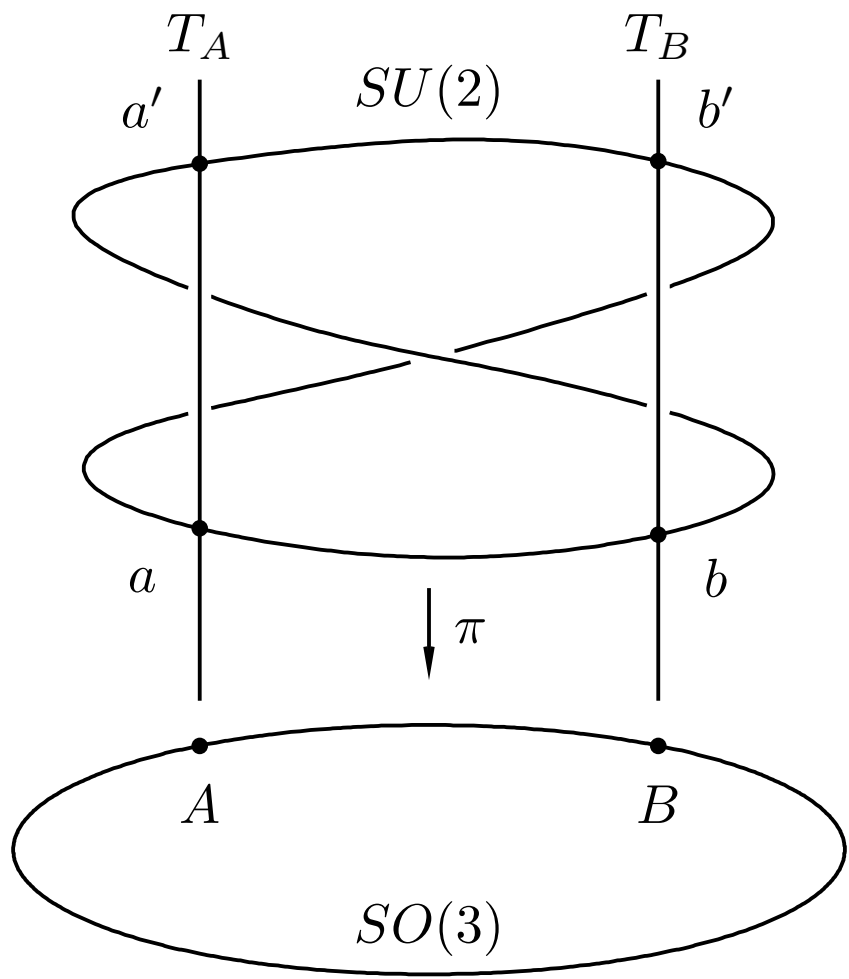
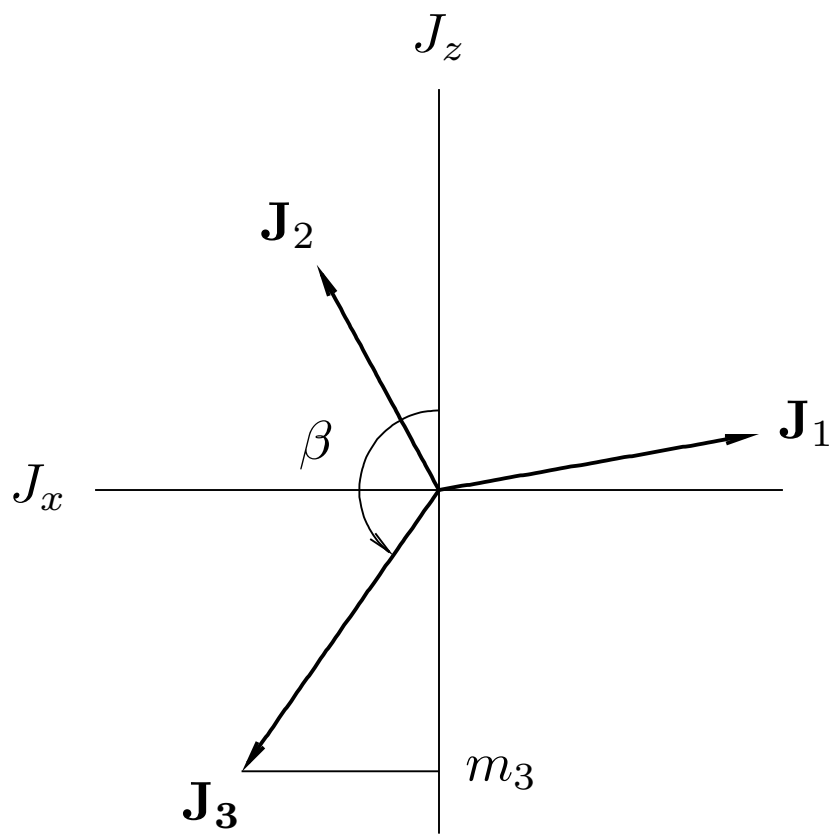
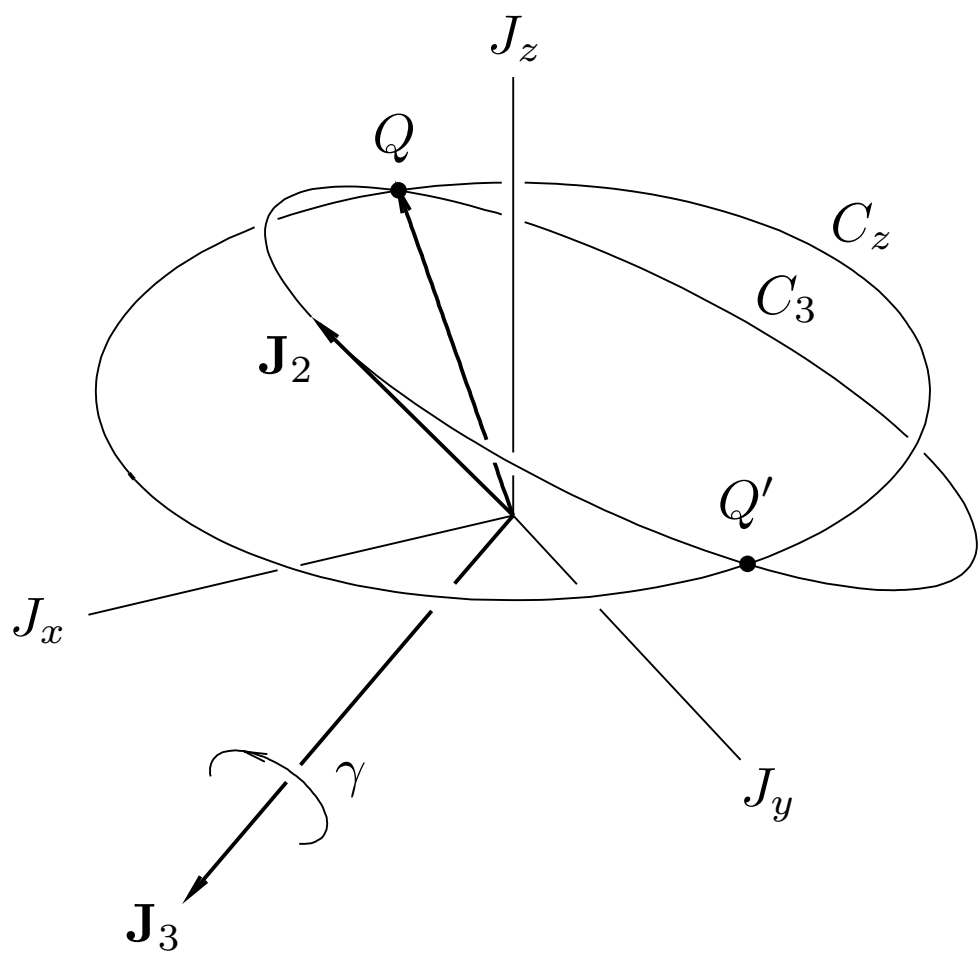
\end{document}